\documentclass[12pt]{article}

\usepackage{epsfig,rotating}   

\newcommand{\degc}{$^\circ$C}
\newcommand{\wpm}{Wm$^{-2}$}
\newcommand{\cft}{$^{14}$C}
\newcommand{\bten}{$^{10}$Be}
\newcommand{\cotwo}{CO$_{2}$}
\newcommand{\lappeq}{\mathrel{\rlap{\raise.5ex\hbox{$<$}}{\lower.5ex\hbox{$\sim$}}}}
\newcommand{\gappeq}{\mathrel{\rlap{\raise.5ex\hbox{$>$}}{\lower.5ex\hbox{$\sim$}}}}

\begin{document}           

\pagestyle{empty}

\begin{titlepage}

\begin{center} EUROPEAN ORGANIZATION FOR NUCLEAR RESEARCH 
\end{center}

{\small
\begin{tabbing}
 \=  \hspace{117mm}  \=  \kill 
 \>   \>CERN-OPEN-2001-028 \\ 
 \> \>February 24, 1998 \\[10mm]
\end{tabbing}  }

\begin{center}
\textbf{\large{Beam Measurements of a CLOUD \\  (Cosmics Leaving
OUtdoor Droplets) \\ Chamber}} \\[6ex]

\normalsize{Jasper Kirkby  \\ CERN, Geneva, Switzerland} \\[6ex]
\end{center}

\begin{abstract} 
\noindent
 A striking correlation has recently been observed  between global cloud
cover and the flux of incident cosmic rays.  The effect of natural variations 
in the cosmic ray flux is large, causing estimated changes in the Earth's
energy radiation balance that are comparable to those  attributed to
greenhouse gases from the burning of fossil fuels since the Industrial
Revolution.  However a direct link between cosmic rays and cloud formation
has not been unambiguously established.   We therefore propose to
experimentally measure cloud (water droplet) formation under controlled
conditions in a test beam at CERN with a CLOUD chamber, duplicating the
conditions prevailing in the troposphere.  These data, which have never
been previously obtained, will allow a detailed understanding of the
possible effects of cosmic rays on clouds and confirm, or otherwise, a direct
link between cosmic rays, global cloud cover and the Earth's climate.   The
measurements will, in turn, allow more reliable calculations to be made of
the residual effect on global temperatures of the burning of fossil fuels, an
issue of profound importance to society.  Furthermore, light radio-isotope
records indicate a correlation has existed between global climate and the
cosmic ray flux extending back over the present inter-glacial and perhaps
earlier.   This  suggests it may eventually become possible to make
long-term (10--1,000 year) predictions of changes in the Earth's climate,
provided a deeper understanding can be achieved of the ``geomagnetic
climate'' of the Sun and Earth that modulates the cosmic-ray flux.
\end{abstract}

\vspace{20mm}


\end{titlepage}

\pagestyle{plain}     

\pagenumbering{roman}  
\setcounter{page}{2}  
\newpage \tableofcontents 

 
\newpage
\pagestyle{plain}     
\pagenumbering{arabic}  
\setcounter{page}{1}  

\section{Introduction} 

\begin{figure}[t]
  \begin{center}
      \makebox{\epsfig{file=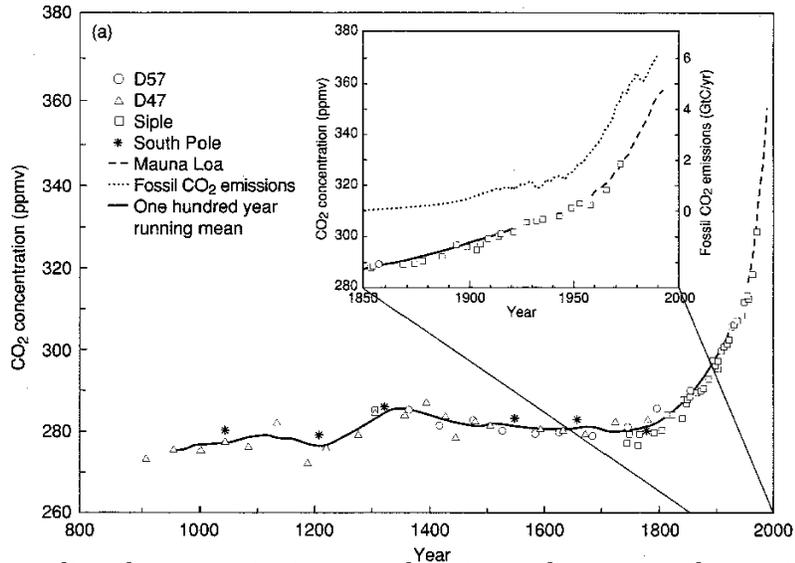,height=75mm}}
  \end{center}
 \vspace*{-10mm}
  \caption{Carbon dioxide concentration in the atmosphere over the past
1000 years \cite{ipcc}.  The data are based on Antarctica ice core records
and, since 1958, on direct measurements from Mauna Loa, Hawaii.  Also
shown are the estimated yearly emissions of CO$_2$ from burning fossil
fuels since the Industrial Revolution.}
  \label{fig_co2}   
\end{figure}

\begin{figure}
  \begin{center}
      \makebox{\epsfig{file=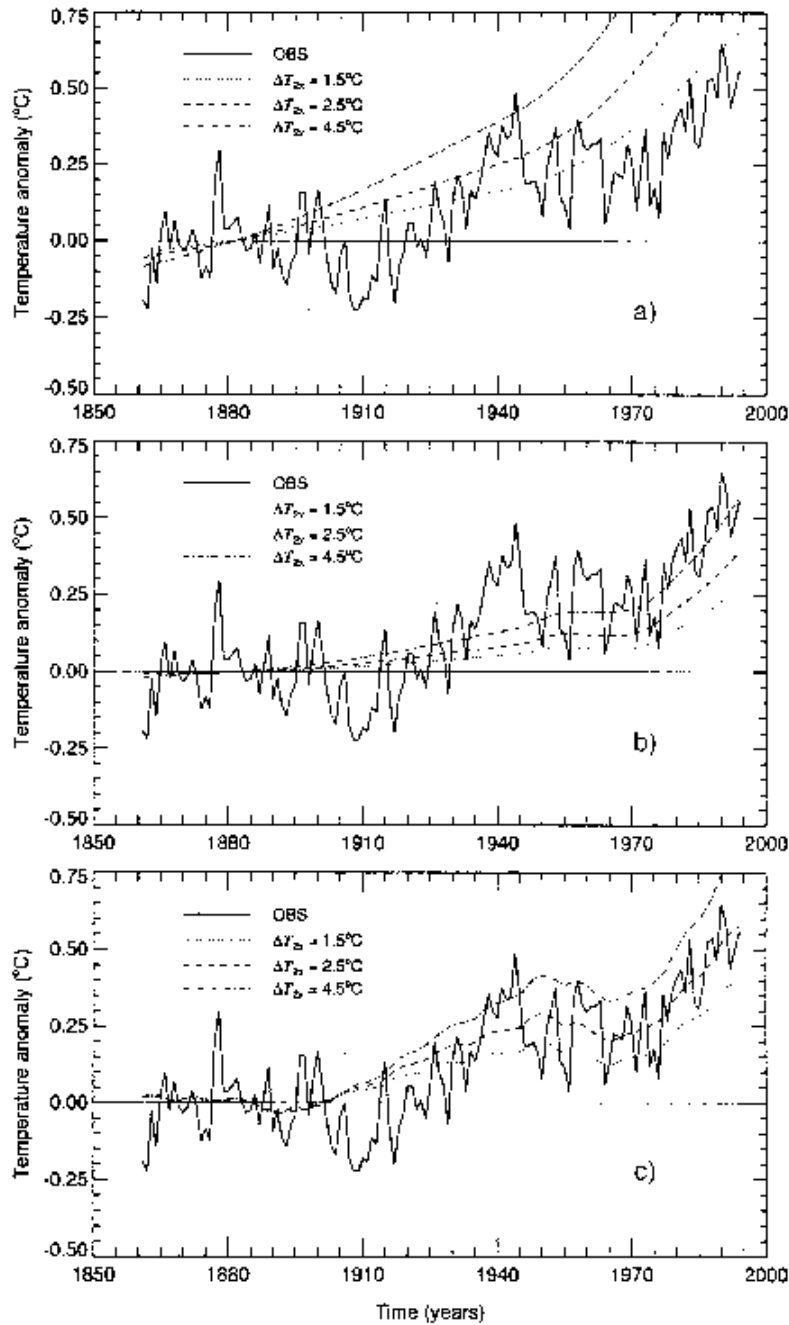,height=175mm}}
  \end{center}
  \caption{Observed changes in global mean surface temperature from 1861
to 1994, compared with the climate model of the  Intergovernmental Panel
on Climate Control (IPCC) on  the effects of greenhouse gas emissions
\cite{ipcc}.  The model was run with a) greenhouse gases alone, b)
greenhouse gases and aerosols, and c) greenhouse gases, aerosols and an
estimate of changes in solar irradiance.  The broken curves are the model
predictions under several different assumptions of the climate sensitivity,
$\Delta T_{2\times}$, which is the change in global mean temperature for a
doubling of the atmospheric \cotwo.}
  \label{fig_temp_vs_time}     
\end{figure}

Global warming is a major concern of the world, with its potentially
devastating effects on coastal settlements and world agriculture.   The steep
rise in greenhouse gas emissions since the Industrial Revolution has
increased the \cotwo\ concentration in the atmosphere by 30\% (Fig.
\ref{fig_co2}). This is widely believed to be the dominant cause of the
observed rise of about 0.6\degc\  in the  global mean surface temperature
during this period  (Fig.
\ref{fig_temp_vs_time}) \cite{ipcc}. 

\begin{figure}
  \begin{center}
      \makebox{\epsfig{file=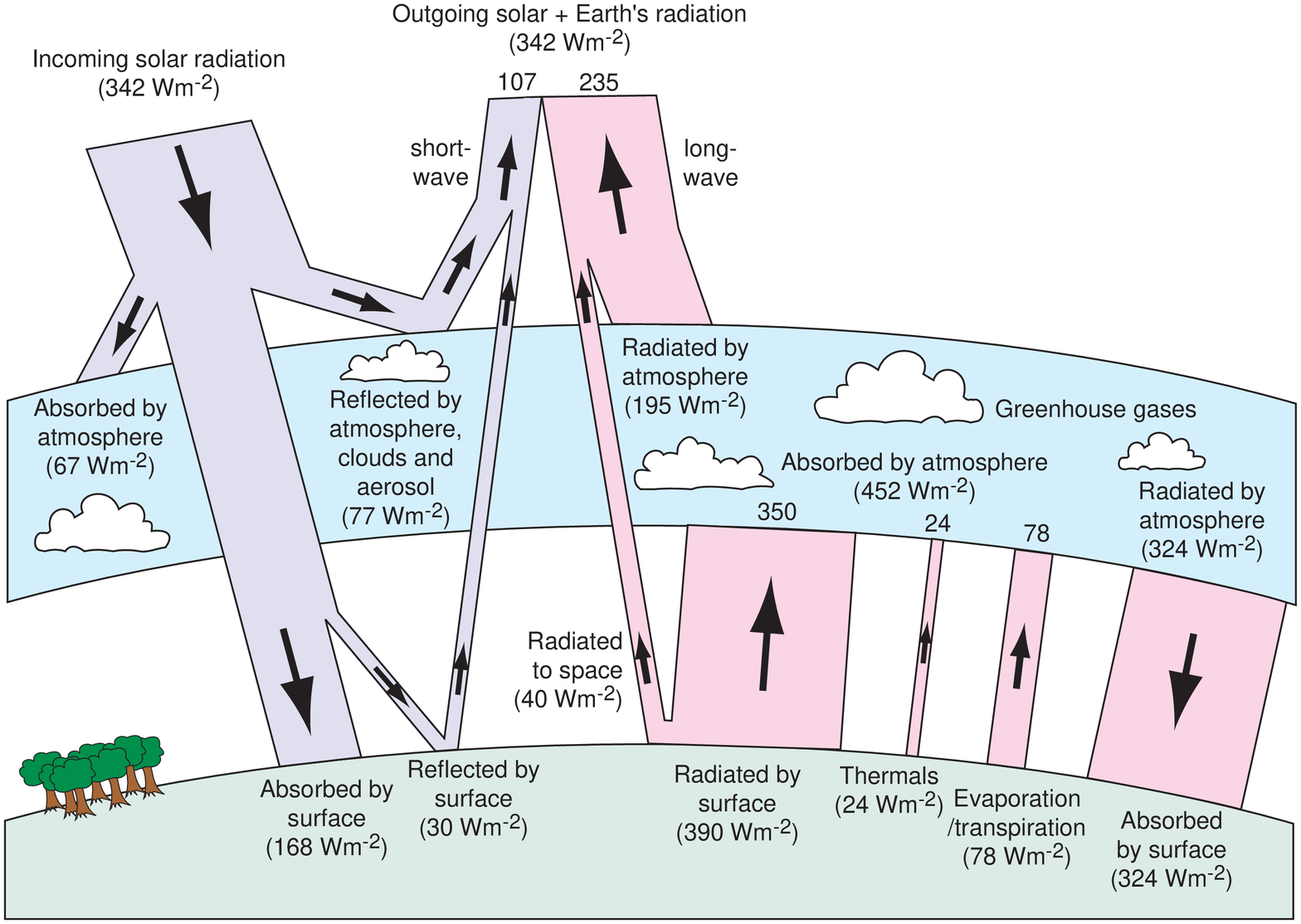,width=140mm}}
  \end{center}
  \caption{The Earth's energy radiation budget.  The width of each band is
proportional to the associated energy flux, which is computed as a global
annual mean. Heat absorbed by the atmosphere produces a greenhouse
effect since the radiation lost to space comes from the cold tops of clouds
and from parts of the atmosphere much colder than the surface. The figure
is adapted from reference \cite{ipcc}.}
  \label{fig_energy_budget}    
  \begin{center}
      \makebox{\epsfig{file=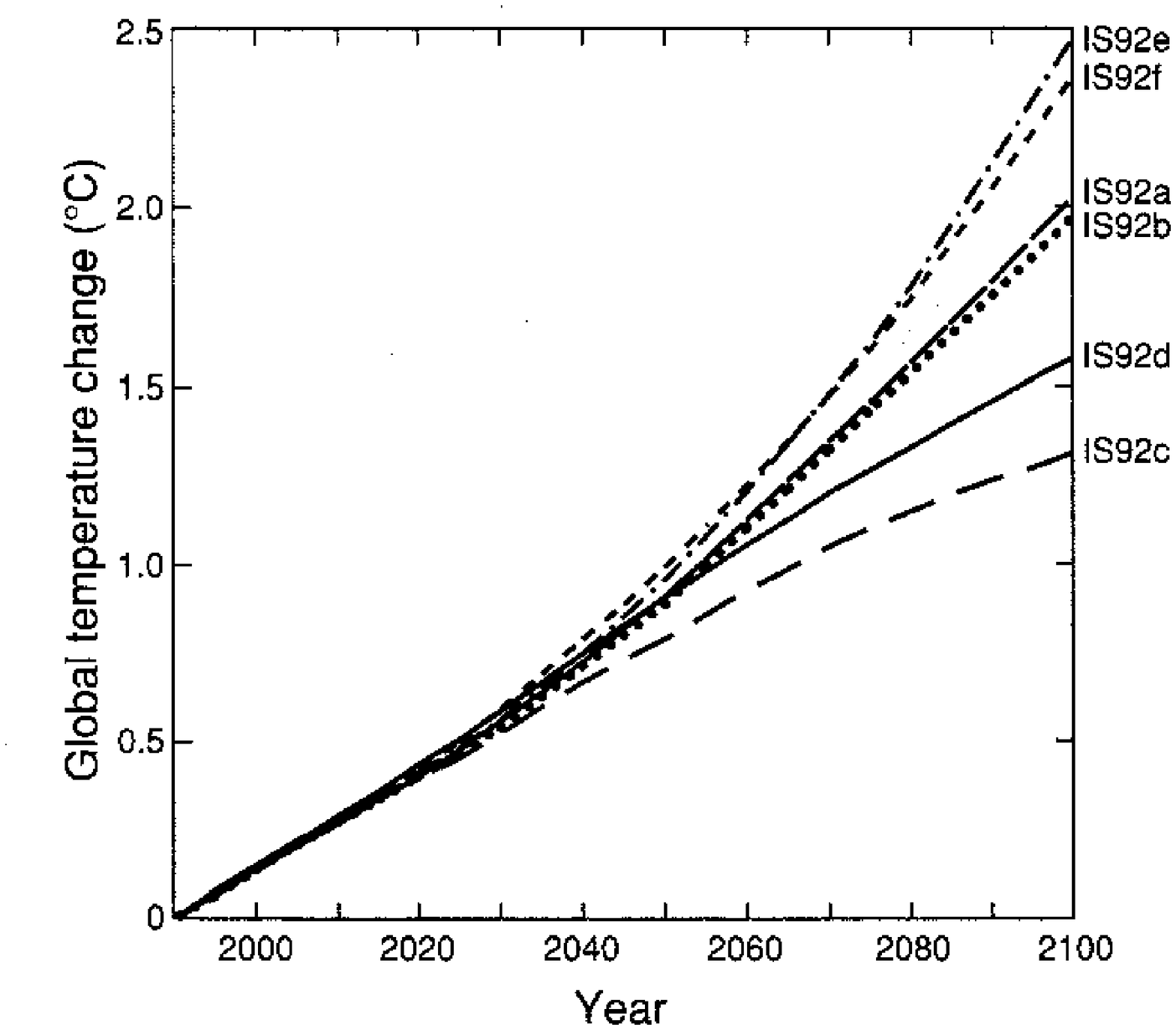,height=67mm}}
  \end{center}
  \vspace*{-5mm}
  \caption{Projected global mean surface temperatures from 1990 to 2100
based on the current IPCC climate model 
 (Fig. \ref{fig_temp_vs_time}c) for the full set of IS92 \cotwo\ emission
scenarios \cite{ipcc}.  A climate sensitivity, 
 $\Delta T_{2\times}$ = 2.5\degc, is assumed.}
  \label{fig_temp_projection}    
\end{figure}

A small systematic rise or fall in the global temperature is caused by a net
imbalance (``forcing'') in the Earth's energy radiation budget  (Fig.
\ref{fig_energy_budget}).  The radiative forcing caused by the increase in
the \cotwo\ fraction since 1750 is estimated to be 1.5
\wpm\ \cite{ipcc}, compared with the global average incoming solar
radiation of 342 \wpm, i.e. an imbalance of only 0.4\%.   After including the
effects of all greenhouse gases (+2.45 \wpm),  aerosols\footnote{Aerosols
are 0.001--1 $\mu$m diameter particles of liquid or solid in suspension. 
Atmospheric aerosols include dust, sea salt, soot (elemental carbon), organic
compounds from biomass burning, sulphates (especially
(NH$_4$)$_2$SO$_4$) from SO$_2$, and nitrates from NO and NO$_2$. 
Aerosol concentrations vary from a few per cm$^{3}$ in clean maritime air
to $10^6 $ cm$^{-3}$ in highly-polluted city air.} 
\mbox{(-0.5~\wpm)} and their effects on clouds (-0.75 \wpm), the present
net radiative forcing from mankind is estimated to be 1.2
\wpm. 

The climate models \cite{ipcc}  upon which the predictions of greenhouse
warming depend have gradually improved as new effects and better data
have been incorporated.  They now provide a reasonable representation of
the observed variation in global temperature over the last century   (Fig.
\ref{fig_temp_vs_time}).  However they remain subject to  significant
uncertainties, both from  unknown processes and also from the unknown
effects of feedback mechanisms which may amplify or damp a warming
trend. Important unknown effects include those due to increased
evaporation of water (which may cool due to an increased cloudiness or
warm due to an increased greenhouse effect),  shifts in the amount of
CO$_2$ dissolved in the oceans, changes in the ocean currents and changes
in the polar ice.   Given the uncertainties in these models, many scientists
consider there is no convincing evidence at present to demonstrate  that a
greenhouse warming of the planet is in progress, let alone that the models
can reliably predict future temperature changes (the predicted rise by the
end of the next century is 1.3--2.5\degc; Fig. \ref{fig_temp_projection}).   

Nonetheless it is in this climate of scientific uncertainty that major political
decisions on greenhouse gas emissions are presently being made (Earth
Summit in Rio de Janeiro, 1992, and UN Climate Convention in Kyoto, 1997)
that will have a profound effect on the economic development of both the
developed and the developing countries.  The need for such political
decisions to be based on sound scientific grounds is self evident, and a
major world-wide research effort on climate change is underway.

An unexpected and important factor in climate change  has recently been
discovered: cloud cover seems to follow natural variations in the incident
cosmic ray flux \cite{svensmark97,svensmark98}.  However a direct link
between cosmic rays and cloud formation has not been unambiguously
established and, moreover, the microphysical mechanism is poorly
understood.   This presents particle physics with a unique opportunity to
make a major contribution to the problem of global warming by confirming,
or otherwise, this link under controlled conditions in a test beam.   The
purpose of this paper is to provide an initial outline of the motivation and
conceptual design of the proposed CLOUD  (Cosmics Leaving OUtdoor
Droplets) detector. 

\section{The origins of climate change}

The Earth's climate has not always been the same; significant changes have
occurred during recorded history and even greater changes have taken
place since the human ancestral line began about 5 million years ago.  The
urgent questions facing science are why these natural changes took place
and whether there is currently an influence of mankind on the climate. 
Only when the natural causes are understood and properly accounted for in
climate models can we expect meaningful results on the anthropogenic
influences. 

Several natural effects are well established.  These include the Milankovitch
cycles---variations in the Earth's orbit, tilt and spin with respect to the
Sun---which seem to control the timing of glacials and interglacials.  These
cold and warm phases correspond to cycles of about 100,000 years and
10,000 years, respectively. (We are presently approaching the end of an
interglacial; sometime in the next 1000 years or so the climate will
probably become much colder.)   Other important natural sources of climate
change are the anomalous warm Pacific current known as El Ni\~{n}o, and
volcanoes that pour ash into the stratosphere, reducing the radiation from
the Sun reaching the Earth's  surface.  A recent volcanic example is Mt.
Pinatubo in the Philippines, which erupted in June 1991 and caused a global
cooling, by up to 0.4\degc, over a period of about 3 years. In contrast, a
natural effect that has until recently been harder to understand is the
apparent link between solar activity---the sunspot\footnote{Sunspots are
areas of the Sun's photosphere where strong local magnetic fields (typically
2500 gauss, to be compared with the Earth's field of about 0.3 gauss)
emerge vertically  \cite{foukal}.  They appear dark because their
temperature is about half of the surrounding photosphere (3,000 K
compared with 5,800 K). They are thought to be generated by the different 
rotation rates of regions of the Sun: one revolution takes 25~days at the
equator and 28 days at mid-latitudes.  This modifies the normal convective
motions of the plasma and creates ``knots'' of strong magnetic field which
penetrate the photosphere and provide escape paths for the plasma,
forming sunspots.}  cycle---and the weather.

\subsection{Activity of the Sun}

\begin{figure}[t]
  \begin{center}
      \makebox{\epsfig{file=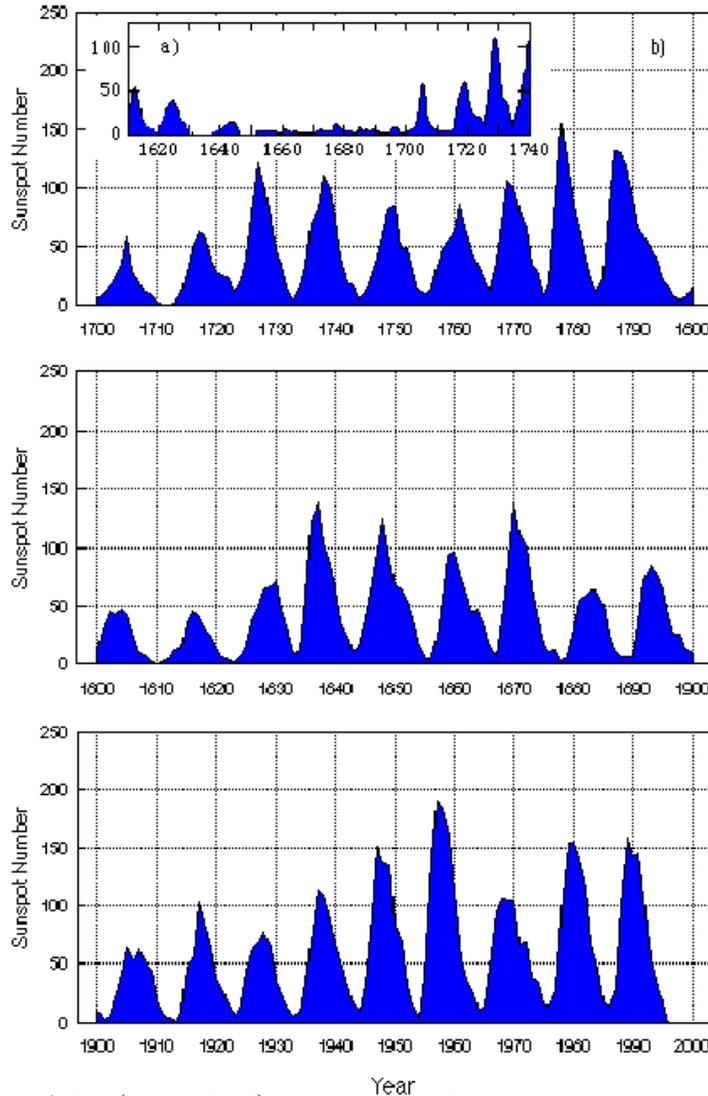,height=145mm}}
  \end{center}
   \vspace*{-10mm}
  \caption{Variation of the (smoothed) sunspot number since systematic
(telescope) observations began.  The data show a) the period 1610--1740
\cite{eddy} and b) the period 1700--1995 [the data are from NOAA---the
National Oceanic and Atmospheric Administration].}
  \label{fig_sunspots_vs_time}    
\end{figure}

The observation that warm weather seems to coincide with high sunspot
counts and cool weather with low sunspot counts was made as long ago as
two hundred years by the astronomer William Herschel who noticed that
the price of wheat in England was lower when there were many sunspots,
and higher when there were few.   The most well-known example of this
effect is known as the Maunder Minimum \cite{eddy}, the Little Ice Age
between 1645 and 1715---which ironically almost exactly coincides with
the reign of Louis XIV, \emph{le Roi Soleil}, 1643--1715---during which
time there was an  almost complete absence of sunspots (Fig.
\ref{fig_sunspots_vs_time}a).  During this period the River Thames in
London regularly froze across and fairs complete with swings, sideshows
and food stalls were a standard winter feature.

\begin{figure}
  \begin{center}
      \makebox{\epsfig{file=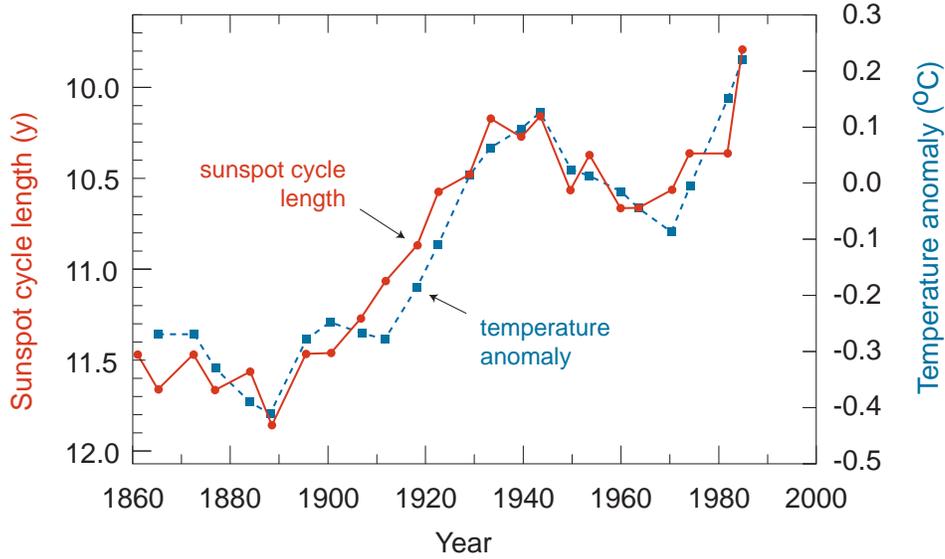,height=75mm}}
  \end{center}
  \caption{Variation during the period 1861--1989 of the sunspot cycle
length (solid curve) and the temperature anomaly of the Northern
Hemisphere (dashed curve) \cite{friis}.  The temperature data from the
IPCC \cite{ipcc}.}
  \label{fig_sunspots_vs_temp}  
\end{figure}

Quantitative proof of a correlation between the Sun's activity  and the
Earth's temperature was presented by Friis-Christensen and Lassen
\cite{friis} in 1991.  They used the sunspot cycle length as a measure of the
Sun's activity.  The cycle length averages 11 years but has varied from 7 to
17 years (Fig. \ref{fig_sunspots_vs_time}), with shorter cycle lengths
corresponding to a more magnetically-active Sun.  A remarkably close
agreement was found between the sunspot cycle length and the change in
land temperature of the Northern Hemisphere in the period between 1861
and 1989 (Fig.~\ref{fig_sunspots_vs_temp}).  The land temperature of the
Northern Hemisphere was used to avoid the lag by several years of air
temperatures over the oceans, due to their large heat capacity.  This figure
covers the period during which greenhouse gas emissions are presumed to
have caused 
 a global warming of about 0.6\degc .  Two features are of particular note:
firstly the dip between 1945 and 1970, which cannot be explained by the
steadily rising greenhouse gas emissions but seems well-matched to a
decrease in the Sun's activity, and secondly the close correspondence
between the two curves over this entire period, which would seem to leave
little room for an additional greenhouse gas effect. 

\begin{figure}[t]
  \begin{center}
      \makebox{\epsfig{file=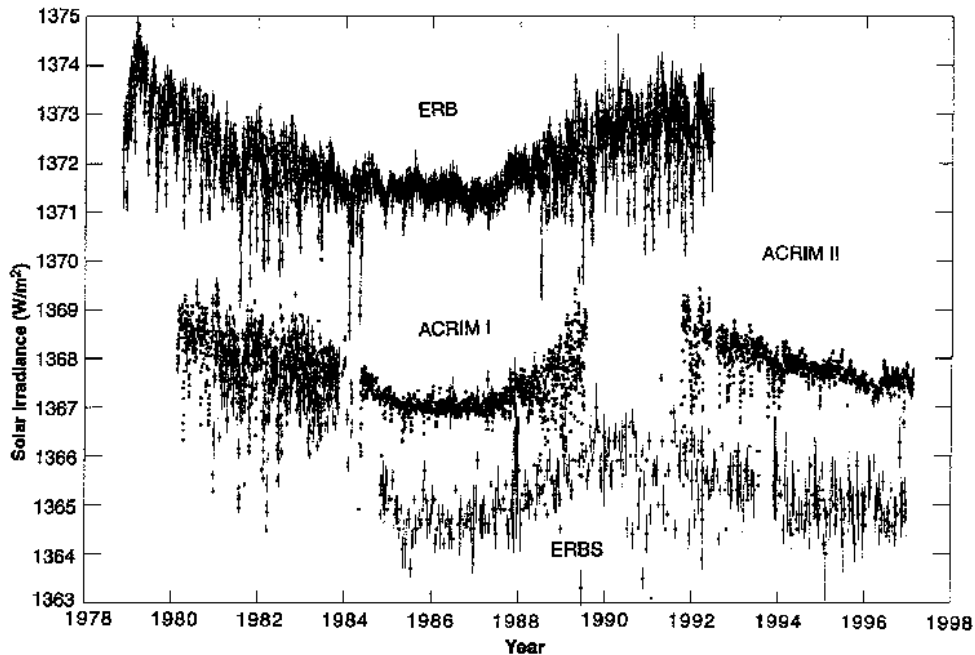,height=86mm}}
  \end{center}
   \vspace*{-7mm}
  \caption{Satellite measurements of the variation of the Sun's irradiance
over a 1.5 sunspot cycles, showing daily mean values and uncertainties
\cite{willson}.  Approximately, sunspot maxima occurred during 1980 and
1990, and sunspot minima occurred during 1985--86 and 1996. The rapid
fluctuations at the maxima are due to sunspots rotating into the field of
view.  The data are from the Active Cavity Irradiance Monitor (ACRIM I
and II) and from the Nimbus 7 Earth Radiation Budget (ERB) and Earth
Radiation Budget Satellite (ERBS) experiments.}
  \label{fig_solar_irradiance}    
\end{figure}

In the absence of sufficiently sensitive measurements, people suspected
that the Sun's irradiance may be fluctuating over the solar cycle.   However,
from our knowledge of how the Sun produces energy, this would seem
difficult.  Radiation from the Sun was created in the core and, in a random
walk with a mean free path of about 10 cm, has taken about fifty million
years to reach the surface.  This would smooth out any substantial
fluctuations in radiated energy on a timescale significantly less than about a
million years.   Indeed the steadiness of the Sun's irradiance over a
complete sunspot cycle has recently been confirmed by satellite
measurements (Fig. \ref{fig_solar_irradiance}
\cite{willson}).  The solar flux is slightly higher at sunspot maximum;
although sunspots are cooler and have reduced emission, this is more than
compensated by an associated increase in bright areas known as
\emph{plages} and \emph{faculae}.   The mean irradiance changes by about
0.1 \% from sunspot maximum to minimum which, if representative over a
longer time interval, is too small  (0.3 \wpm, globally-averaged) to account
for the observed changes in the Earth's temperature.

\begin{figure}[htbp]
  \begin{center}
      \makebox{\epsfig{file=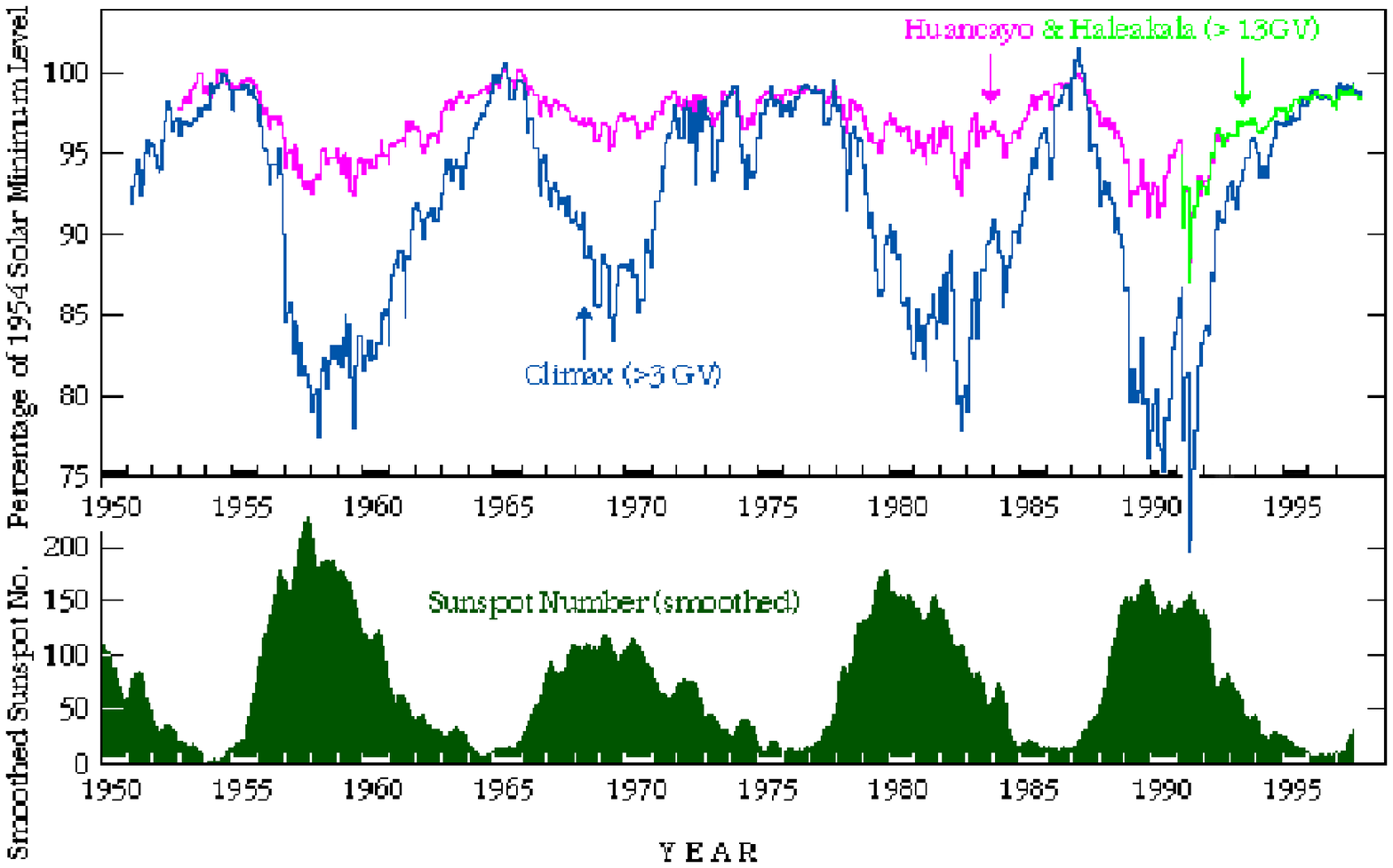,height=82mm}}
  \end{center}
  \vspace*{-8mm}
  \caption{Variation with time of sunspot number and cosmic ray flux, as
measured by ground-based neutron counters.  The neutron data are from
 the Univ. Chicago Neutron Monitor Stations at Climax, Colorado (3400 m
elevation; 3 GeV primary charged particle cutoff), Huancayo, Peru (3400 m;
13 GeV cutoff) and Haleakala, Hawaii (3030 m; 13 GeV cutoff).  The stronger
modulation of the cosmic ray flux at higher latitudes (Climax) is due to the
lower primary cutoff energy.}
  \label{fig_cosmics_vs_sunspots}  
  \begin{center}
      \makebox{\epsfig{file=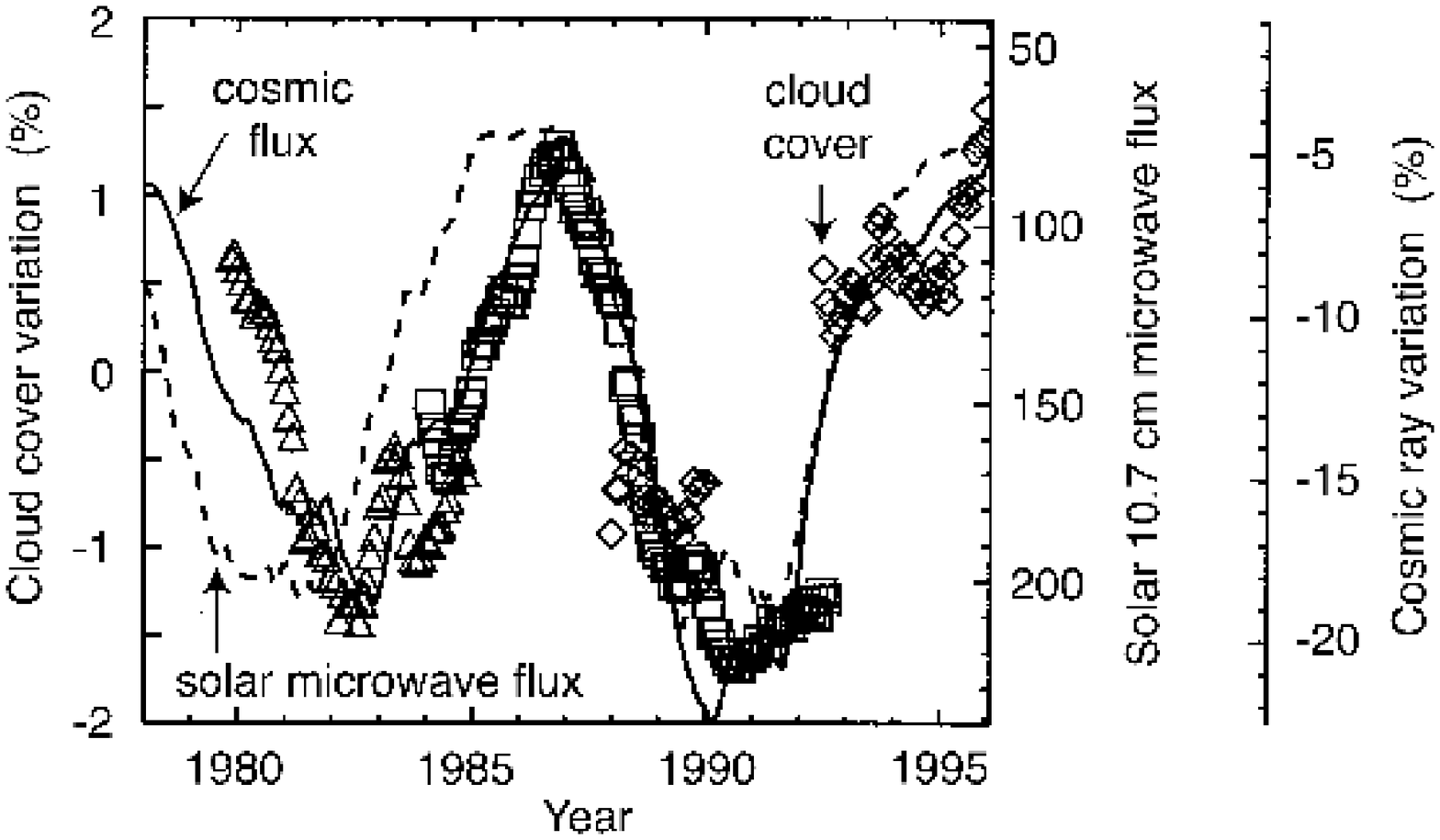,height=71mm}}
  \end{center}
    \vspace*{-8mm}
  \caption{Absolute percentage variation of global cloud cover (data points;
left-hand scale) and  relative percentage variation of cosmic ray flux (solid
curve, normalised to May 1965; right-hand scale)
\cite{svensmark97,svensmark98}.  Also shown is the solar 10.7 cm
microwave flux (dashed curve, in units of $10^{-22}$ Wm$^{-2}$Hz$^{-1}$;
right-hand scale).   The cloud data are restricted to oceans; Nimbus 7
(triangles) and DMSP (diamonds) data are for the total Southern
Hemisphere over oceans, and  ISCCP (squares) data are for oceans with the
tropics excluded. The cosmic ray data are neutron measurements from
Climax (see Fig.    \ref{fig_cosmics_vs_sunspots}). All data are smoothed
using a 12-month running mean.}
  \label{fig_clouds_vs_cosmics}    
\end{figure}

\subsection{The missing link: cosmic rays?} \label{sec_link}

\paragraph{Experimental observations.}

It has been recognised for several decades that the cosmic ray flux reaching
the Earth is strongly modulated by the solar wind\footnote{The solar wind
is a continuous outward flow of charged particles  (mainly protons and
electrons, with 5\% helium nuclei) from the plasma of the Sun's corona.  The
main sources are near the Sun's poles; other important sources include
coronal holes and sunspots.   The solar wind creates the huge heliosphere of
the Sun that extends out to Neptune and beyond.  At the Earth's orbit the
solar wind has a velocity of 300--800~km~s$^{-1}$ ($\beta = $
0.001--0.003) and an intensity of
\mbox{(0.5--5)$ \times 10^8$ particles~cm$^{-2}$~s$^{-1}$}, generating a
magnetic field of about  $5 \times 10^{-5}$ Gauss.} which, in turn, is
strongly influenced by the sunspot cycle (Fig.
\ref{fig_cosmics_vs_sunspots}). At times of high sunspot  activity, the solar
wind is stronger and this partially shields the lower-energy galactic cosmic
radiation from  penetrating the inner solar system and reaching the Earth.  
In addition, the lower-energy cosmic rays are affected by the geomagnetic
field, which they must penetrate to reach the top of the atmosphere.  The
minimum vertical momentum of primary charged particles at the
geomagnetic equator is about 15 GeV/c, decreasing to below 0.1 GeV/c at
the geomagnetic poles.  The modulation of the cosmic ray flux incident on
the Earth's atmosphere at the higher geomagnetic latitudes is as large as
50\% between sunspot maxima and minima
\cite{ney}.  It represents one of the largest measurable effects of sunspot
activity near the Earth's surface.

But how could cosmic rays affect the Earth's weather? The breakthrough
was made by Svensmark and Friis-Christensen in 1997 
\cite{svensmark97,svensmark98} who showed a striking correlation
between global cloud cover and the incident cosmic ray flux  (Fig.
\ref{fig_clouds_vs_cosmics}).  Moreover, the cloud cover was found to have
a poorer correlation with variations in the solar microwave flux
(Fig.~\ref{fig_clouds_vs_cosmics}).  The latter is known to follow closely
variations in the total solar irradiance, soft X-rays and ultraviolet rays. The
modulation in cloud cover was found to be more pronounced at higher
geomagnetic latitudes, consistent with the reduced shielding effect of the
Earth's magnetic field at higher latitudes (see Fig.
\ref{fig_cosmics_vs_sunspots}). 

Over a sunspot cycle, the absolute variation in global cloud cover is about
3\% (equivalent to 4.8\% relative fraction), and the variation in the neutron
flux is about 15--20\% (for a primary charged particle cutoff of 3 GeV).  The
neutrons are mostly produced by primary hadronic interactions in the first
1--2
$\lambda_{int}$ of the atmosphere and therefore measure the changes in
cosmic ray intensity at altitudes above about  13 km.   The primary cosmic
radiation is about 80\% protons, 15\% He nuclei and 5\% heavier nuclei.   At
sea level the most numerous
\emph{charged} particles are muons and their fluctuation is less
pronounced, about 3\%  over a solar cycle \cite{ney}, since they are
produced from a stiffer primary spectrum of cosmic radiation, which is
therefore less affected by the solar wind.

\paragraph{Fraction of clouds affected by cosmic rays.}
 
These data suggest that cosmic radiation is, under suitable atmospheric
conditions, either extending cloud lifetimes (for example, by increasing the
droplet number density) or  inducing some additional cloud formation. The
currently-accepted physical processes that affect the production of cloud
condensation nuclei  \cite{hobbs,baker}, including both aerosols and ice
particles, do not consider ion-induced effects. The new satellite observations
suggest that  the ions or radicals produced in the atmosphere by cosmic
rays may somehow be affecting the nucleation, growth or activation of
atmospheric aerosols, or the creation of ice particles.

It is interesting to estimate the total cloud fraction that may be caused by
cosmic rays.  It is not yet known at which altitude the cosmic radiation may
be affecting clouds, i.e. whether the clouds more closely follow the
amplitude modulation of the neutron flux (20\%) or of the muon flux (3\%). 
However, making the most conservative assumption, a 20\% change in
cosmic ray flux causes a 4.8\% relative change in cloud cover.  If the
cloud-forming process is directly proportional to cosmic flux then this
suggests about 25\% of the Earth's clouds could be affected by cosmic rays. 
However, if the process is proportional to ion density then the fraction is
twice as large, since a 20\% increase in cosmic ray flux leads to a 10\%
increase in equilibrium ion density (see Section \ref{sec_ions}).  In
summary, a large fraction of the Earth's clouds could be affected by cosmic
radiation.

\paragraph{Effect on the Earth's radiation budget.}

Thin clouds at high and middle altitudes cause a net warming due to an
increased trapping of  outgoing longwave radiation whereas optically thick
clouds produce a net cooling due to a dominant  increase in the albedo
(reflection) of the incoming shortwave solar radiation.  At present there is
no indication which clouds may be affected by cosmic rays, so we will
assume for the present estimate that all are affected equally.  
 Present estimates from the Earth Radiation Budget Experiment (ERBE)
indicate, overall, that clouds reflect more energy than they trap, leading to a
net cooling of 28
\wpm\ from the mean global cloud cover of 63\% (Table \ref{tab_erbe}
\cite{hartmann}).   The 4.8\% relative variation in cloud cover due to
variations of the cosmic ray flux over a solar cycle therefore implies a
change in the Earth's radiation budget of 1.3~\wpm\  (0.4\% of the total
incident solar radiation).   This is a  significant effect: in the 3-year period
1987--1990, the effect of the reduction in cosmic ray flux was comparable
to  the total estimated radiative forcing  of 1.5~\wpm\  from the increase in
CO$_2$ concentration since 1750.

\begin{table}[h]
  \begin{center}
  \caption{Global annual mean forcing due to various types of clouds,  from
the Earth Radiation Budget Experiment (ERBE) \cite{hartmann}. The sign is
defined so that positive forcing increases the net radiation budget of the
Earth and leads to a warming;  negative forcing decreases the net radiation
and  causes a cooling.}
  \label{tab_erbe}
  \vspace{5mm}
  \begin{tabular}{| l  l | r r | r r | r | r |}
  \hline
   Parameter& & \multicolumn{2}{|c|}{High clouds} &
 \multicolumn{2}{|c|}{Middle clouds} &   Low clouds & Total \\
  \cline{3-7}  
   & & Thin & Thick & Thin & Thick & All &  \\
  \hline
  \hline
  &  &  & & & & &  \\[-3ex]
 Global fraction & (\%) & 10.1 & 8.6 & 10.7 & 7.3 & 26.6 & 63.3 \\
\hline
  \multicolumn{2}{|l|}{Forcing (relative to clear sky):}   &  & & & & &  \\
 Albedo (SW radiation) & (\wpm) & -4.1 & -15.6 & -3.7 & -9.9
 & -20.2 & -53.5  \\
 Outgoing LW radiation & (\wpm) & 6.5 & 8.6 & 4.8 & 2.4 & 3.5  &
 25.8 \\
\hline Net forcing & (\wpm) & 2.4 & -7.0 & 1.1 & -7.5 & -16.7 & -27.7
\\[0.5ex]
  \hline
  \end{tabular}
  \end{center}
\end{table}

\paragraph{Open questions.}

Despite the observation of Svensmark and Friis-Christensen
(Fig.~\ref{fig_clouds_vs_cosmics}), a direct link between cosmic rays and
cloud formation has not been unambiguously established.   The satellite
data are a composite of several datasets which are subject to
inter-calibration uncertainties.  Furthermore the data are spatially and
temporally selective - they are restricted to oceans and exclude the tropics
and polar regions, and they are largely daytime only.  

It has been argued that the single sunspot cycle for which there are reliable
satellite measurements is insufficient to demonstrate a clear connection and
that several more cycles (each of course requiring 11 years) are needed.
Even if this extended study were to confirm the correlation between the
sunspot cycle and cloud cover, it may not prove the missing link is cosmic
rays.  While cosmic rays certainly vary with the sunspot cycle, other factors
that change during the cycle may be affecting clouds, quite independently
of cosmic rays. For example, the cause may be variations of the solar
ultra-violet (UV) radiation.  Although the total irradiance of the Sun varies
by only 0.1 \% over the solar cycle, the variation is more pronounced in the
UV.  Wavelengths below 320 nm, which account for only about 2\% of the
total solar irradiance, vary by about 5\% (200--320 nm) to 50\% (100--150
nm), or even more at shorter wavelengths \cite{hunten}.  The UV radiation
is absorbed by ozone high in the stratosphere which warms as a result. 
This has the potential to influence large scale tropospheric dynamics and
hence, perhaps, cloudiness. 

Finally, the evidence of increased variations in cloud production at higher
latitudes, while consistent with the cosmic ray explanation, could simply be
due to the variations with latitude of temperature, air circulation, aerosols
and other atmospheric conditions that affect cloud formation.

\subsection{History of cosmic rays and climate change}
\label{sec_cosmics&climate_change}

If the periodic 11-year cycles of the sunspots and the associated cosmic ray
flux  were the end of the story, then it would be of limited concern since
there would be no resultant long-term change in the Earth's weather but
simply another cyclic ``seasonal'' change (albeit with a period of 11 years).  
However there is clear evidence of longer-term and unexplained changes
both in the Sun and in the Earth's magnetosphere and these, in turn, seem
to have had long-term effects on the Earth's climate. 

\begin{figure}[htbp]
  \begin{center}
      \makebox{\epsfig{file=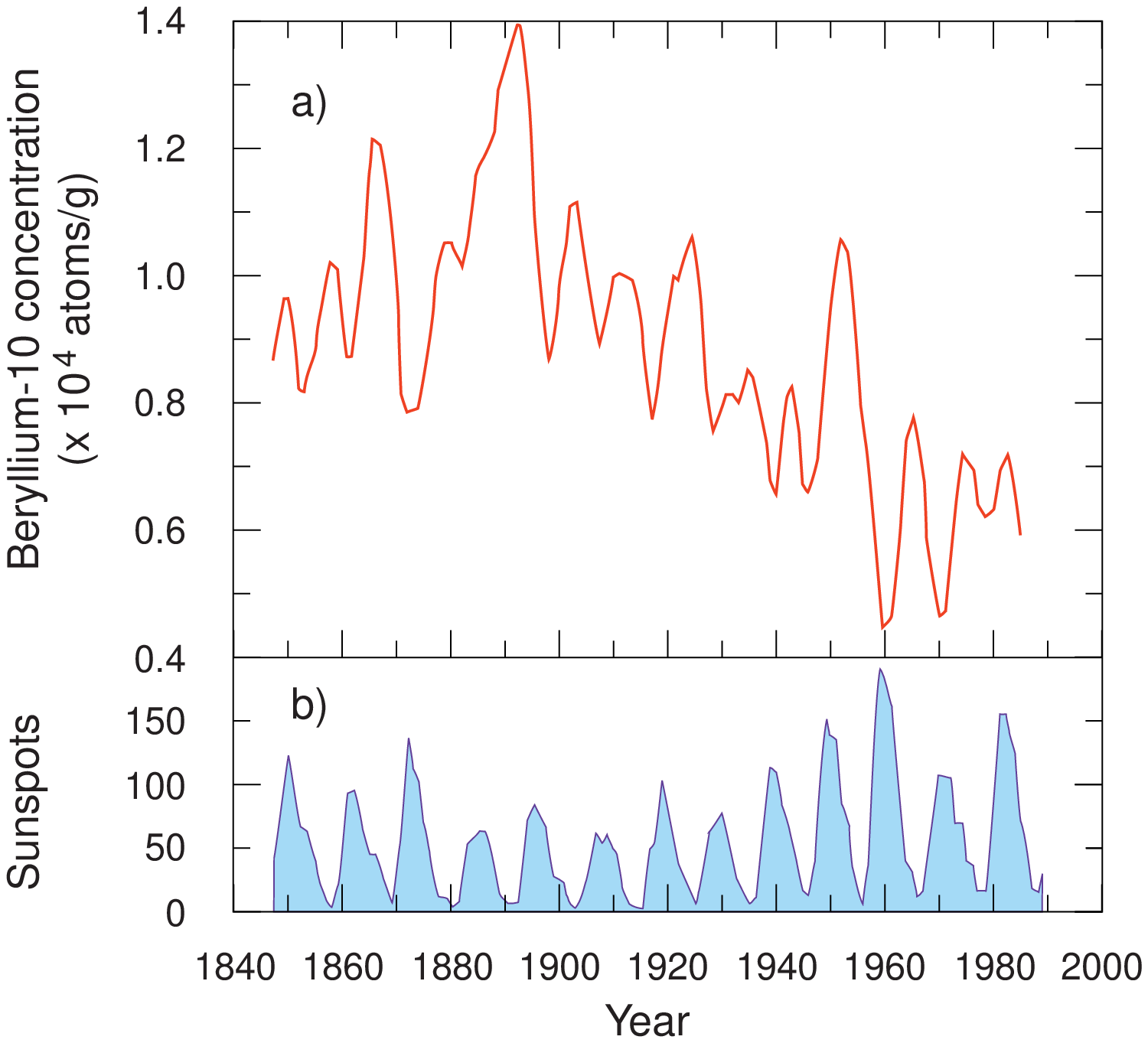,height=78mm}}
  \end{center}
 \vspace*{-8mm}
\caption{a) Concentration of \bten\ in a 300 m ice core from Greenland
over the last 150 years \cite{beer}.  The data are  smoothed by
 an approximately 10 year running mean and have been shifted earlier  by
2~years to account for settling time.  b)~The sunspot cycle over the same
period, which shows a negative correlation with the short-term ($\sim$ 11
year) modulation of the  \bten\  concentration. }
  \label{fig_be10}  
  \begin{center}
      \makebox{\epsfig{file=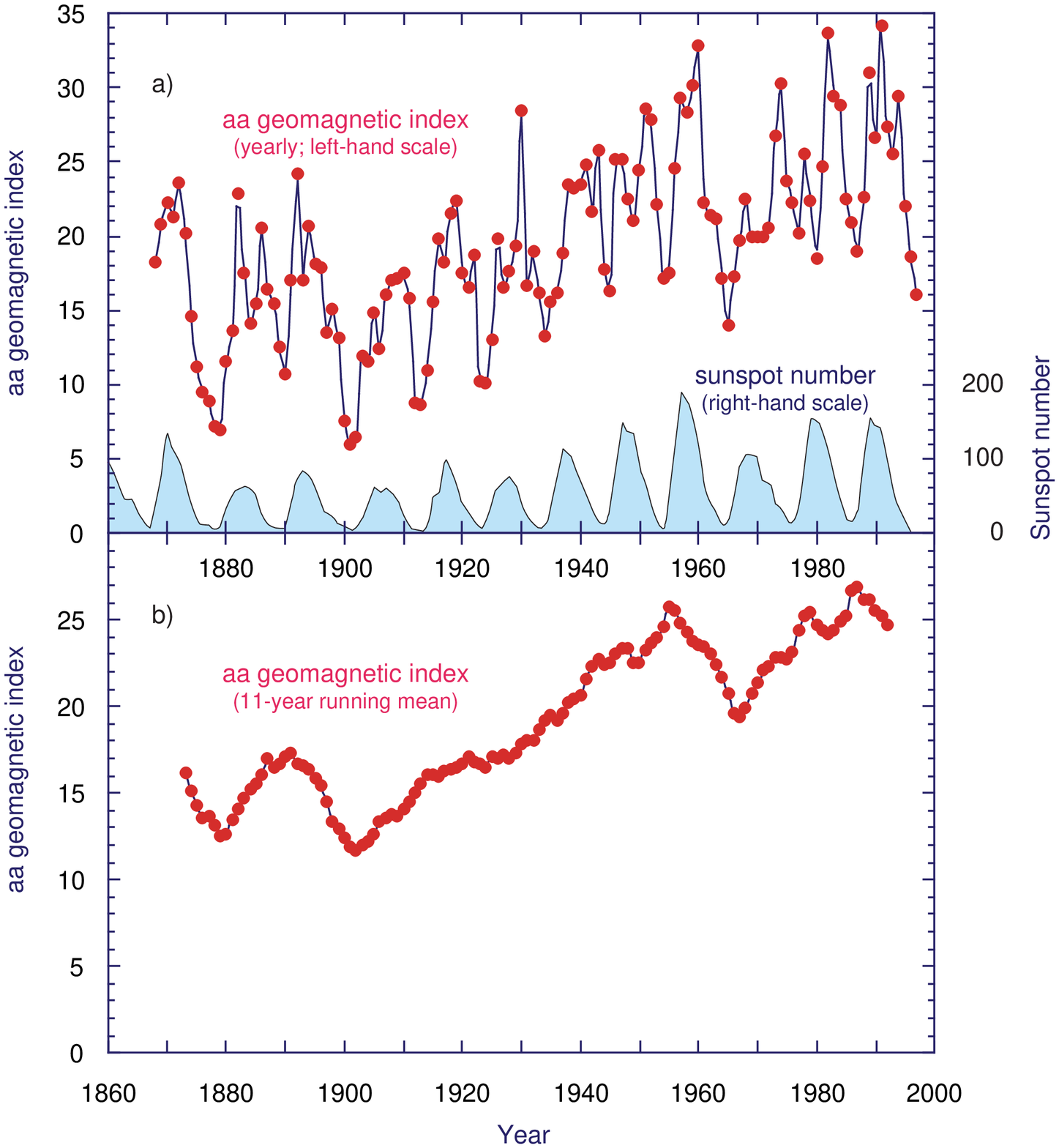,height=103mm}}
  \end{center}
 \vspace*{-8mm}
  \caption{Variation of the aa geomagnetic index over the last 130 years: a)
yearly measurements and b) data smoothed by an 11-year running mean.
The sunspot cycle over the same period is indicated in the upper figure and
shows a positive correlation with the short-term ($\sim$ 11 year)
modulation of the  aa geomagnetic index. }
  \label{fig_geomagnetic_aa_index}    
\end{figure}

\begin{figure}[htbp]
  \begin{center}
      \makebox{\epsfig{file=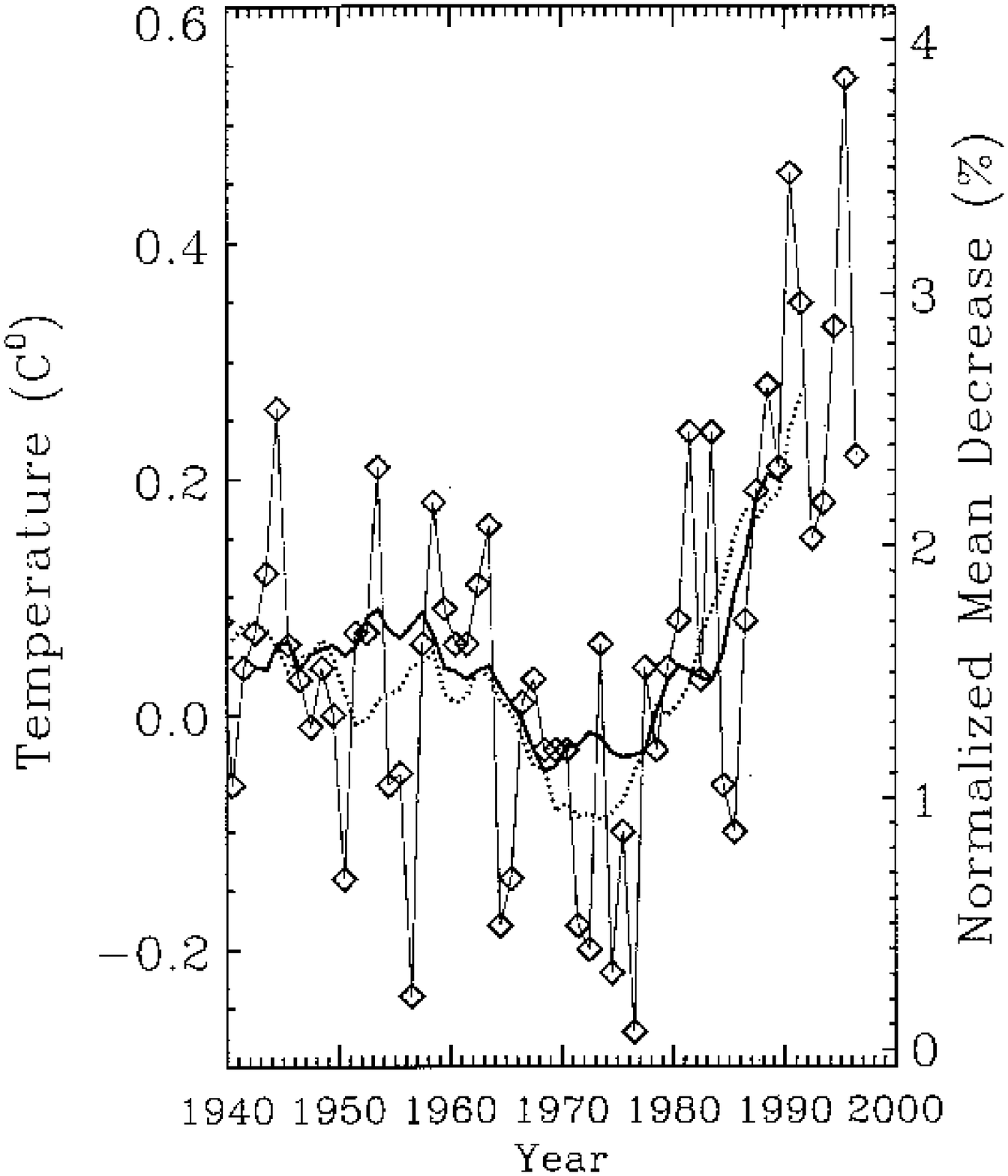,height=83mm}}
  \end{center}
  \vspace{-5mm}
  \caption{Northern hemisphere mean temperature (dotted line; left-hand
scale) and charged cosmic ray flux (thick solid line; right-hand scale) since
1940 \cite{svensmark98}.  Both curves are smoothed by an  11-year
average.  The diamonds show the annual temperatures, before smoothing.}
  \label{fig_cosmics_vs_temp}      
\end{figure}

\subsubsection{Global warming during the present century}
 
Over the course of this century the cosmic ray flux has been steadily
decreasing; it is weaker today at its \emph{maximum} during the sunspot
cycle than it was at its \emph{minimum} around 1900.   This has been
revealed by analysis of the \bten\ concentration in a Greenland ice core
(Fig. \ref{fig_be10} \cite{beer}).  Beryllium-10 is produced in the
atmosphere by cosmic rays and, despite a lower production rate than \cft\
(Section \ref{sec_millenium}) by about a factor 100  ($3.5 \times 10^{-2}$
atoms cm$^{-2}$s$^{-1}$), it has the advantages of settling out relatively
rapidly ($\sim$~2~years) and of a long half-life, $1.51 \times 10^6$ years. 
A systematic  decrease in the cosmic ray flux of the magnitude indicated by
the \bten\ data   would have  caused  a reduction in cloud cover and
consequent warming of the Earth,  sufficient to account for a large part of
the observed rise of 0.6\degc\ in global temperatures over this century
which is presently attributed to anthropogenic greenhouse gases. 

The cause of this systematic decrease of the cosmic flux has been a
strengthening of the solar wind over this century.  This is  revealed in
measurements of the so-called \emph{aa geomagnetic index} of short-term
(up to 3-hour) variations in the geomagnetic field at the Earth's surface
\cite{feynman,stuiver}, which is affected by the interactions of the solar
wind with the Earth's magnetosphere (Fig.
\ref{fig_geomagnetic_aa_index}).     With this hindsight, the previous
observation by Friis-Christensen and Lassen that the sunspot 
\emph{cycle length} is closely  correlated with global temperatures (Fig.
\ref{fig_sunspots_vs_temp}) can now perhaps be understood: the sunspot
cycle length is a direct measure of the strength of the solar wind and a
better measure than the sunspot number itself.  Shorter sunspot cycles
correspond to a stronger solar wind and therefore to decreased cosmic flux,
decreased cloud cover and increased temperatures.

Although a comparison of global temperatures with direct measurements of
cosmic rays can only be made since 1937, when systematic measurements
of charged cosmic rays began with electronic detectors, the two are indeed
found to correlate well  (Fig.
\ref{fig_cosmics_vs_temp} \cite{svensmark98}).  

\subsubsection{Climatic change over the past millennium}
\label{sec_millenium}

The observations of a correlation between the Sun's activity, cosmic rays
and the Earth's climate may be extended to earlier times with
\cft\ data. This isotope is continuously formed in the atmosphere by low
energy neutrons from cosmic radiation at a rate of about 2.5 atoms
cm$^{-2}$s$^{-1}$, in the reaction:
\[  ^{14}\mbox{N + n} \;  \rightarrow  \;  ^{14}\mbox{C + p}  \]  The \cft\  is
rapidly oxidised to form \cft O$_2$ and then decays by
$\beta^-$ emission with a half life of  ($5730\pm40$) years.   The turnover
time of \cotwo\ in the atmosphere is quite short, about 4 years, mostly by
absorption in the oceans and assimilation in living plants.  However,
recirculation from the oceans has a smoothing effect such that changes in
the \cft\ fraction on timescales less than a few decades are smoothed out. 
Plant material originally contains the prevailing atmospheric fraction of
\cft\ and subsequently, since the material is not recycled into the
atmosphere, the fraction decreases with the characteristic half life of
\cft.

\begin{figure}[t]
  \begin{center}
      \makebox{\epsfig{file=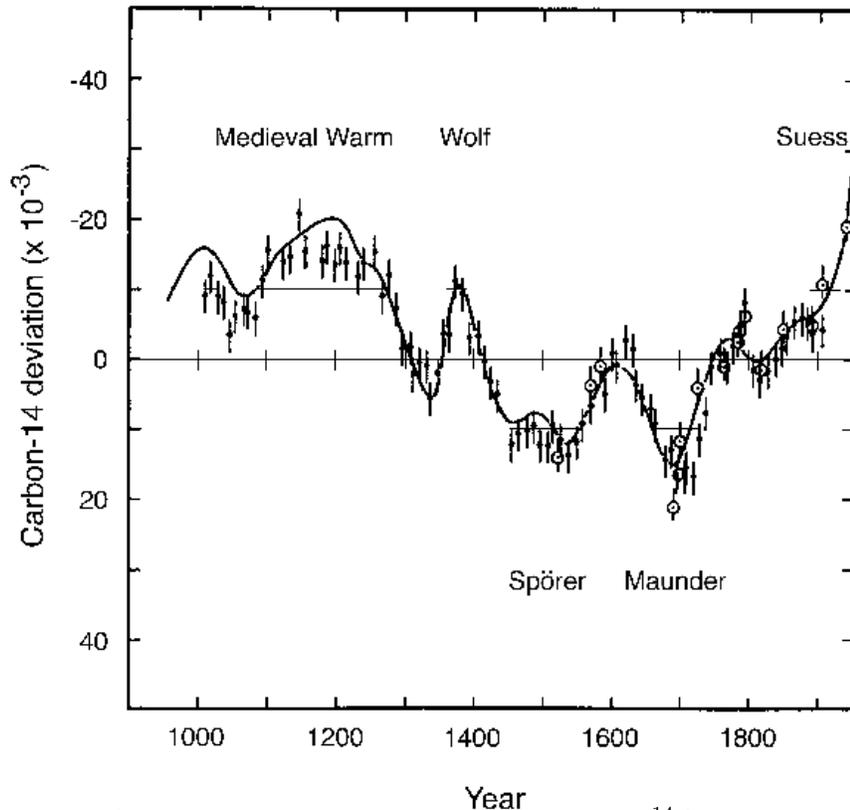,height=107mm}}
  \end{center}
   \vspace{-10mm}
  \caption{History of deviations in the relative atmospheric \cft\
concentration from tree-ring analyses for the last millennium 
\cite{damon}.  The data points (dots and open circles) are two independent
high-precision measurements.  The solid line represents a combined fit to a
large number of other measurements of medium precision.  Periods where
the deviation of the data points exceeds 10 parts per mil are indicated.  The
first four coincide with recorded climatic anomalies.   The sharp negative
\cft\ deviation during the present century is the Suess effect, due to the
burning of \cft-depleted fossil fuels. }
  \label{fig_c14_1k_damon}     
\end{figure}

By analysing the \cft\ content in the rings of long-lived trees such the
Californian bristlecone pine, a year-by-year record has been assembled of
the cosmic ray flux on Earth over the past several thousand years.  The data
for the last 1000 years are shown in Fig.
\ref{fig_c14_1k_damon} \cite{eddy}. (In this the following two figures,   the
vertical axis is oriented so that warmer temperatures point upwards.)  It is
interesting to note that the periods where the \cft\ deviation exceeds 10
parts per mil correspond to recorded climatic anomalies: a)~1100--1250,
the so-called Medieval Warm period,  b) 1300--1360, the Wolf maximum,
c)~1460--1550, the Sp\"{o}rer minimum, and d) 1645--1715, the Maunder
minimum.  The warm period that lasted until about 1400 enabled the
Vikings to colonise Greenland and wine to be made from grapes grown in
England.  It was followed by a period of about 400 years during which the
glaciers advanced and cooler, harsher conditions prevailed in most parts of
the world.  

The Maunder Minimum, when there was an almost complete absence of
sunspots, would have corresponded to a high cosmic ray flux on Earth and
therefore an increased cloudiness.  This provides a consistent explanation
for the exceptionally cold weather of this period. Indeed, in every case the
excess (c--d) or lack (a--b) of \cft\ is consistent with the observation of a
higher cosmic ray flux leading to more clouds and cooler temperatures, and
vice versa.  

These data provide evidence for significant temperature anomalies over the
last millennium that coincided with variations of the cosmic ray intensity. 
Furthermore they show that the Earth has experienced several extended
warm and cold spells over the last 1000 years---with climate swings
comparable to the current ``anomaly''---whose origin certainly cannot be
due to anthropogenic greenhouse gases.  Whatever caused those earlier
spells could well be at work today.

\begin{figure}[htbp]
  \begin{center}
      \makebox{\epsfig{file=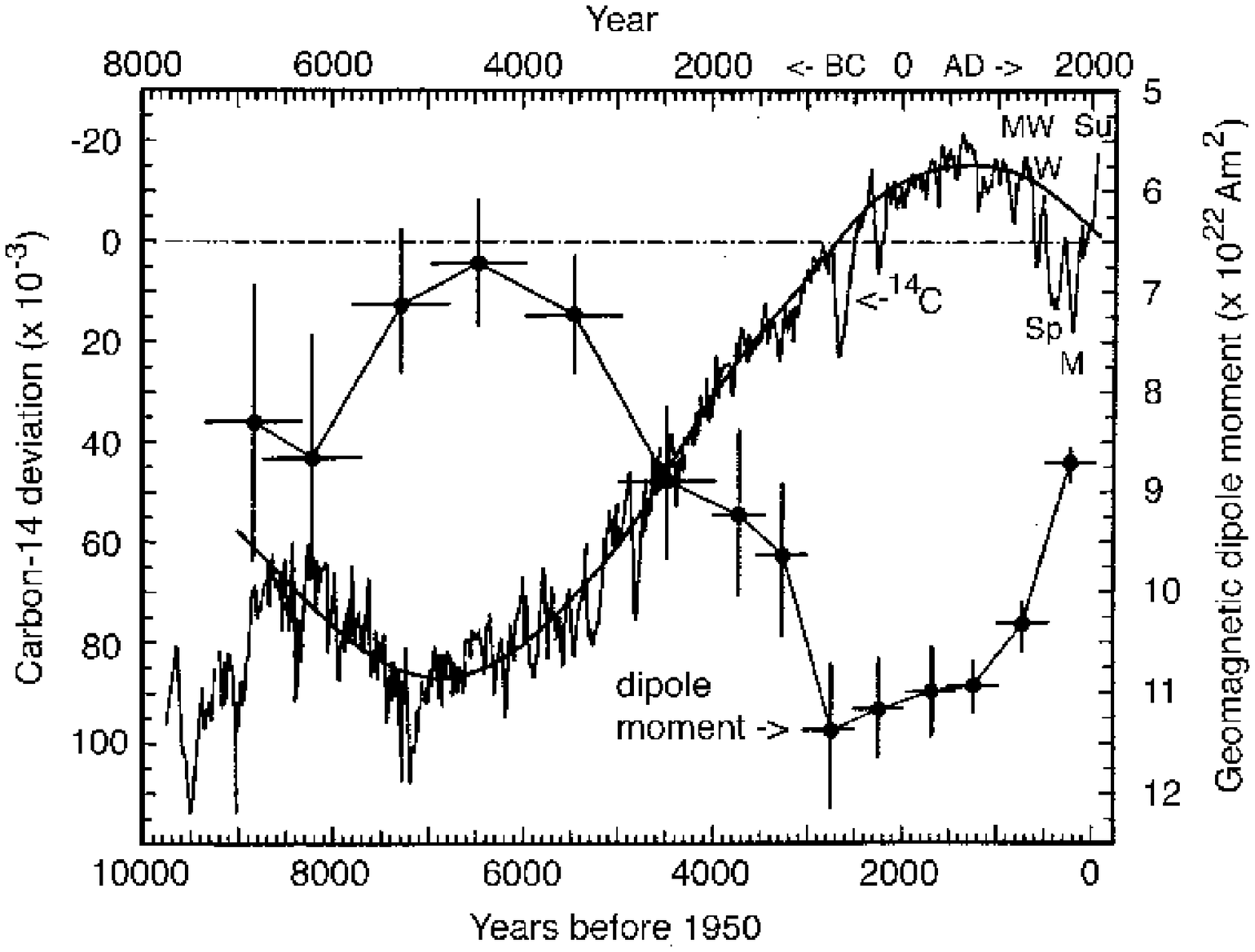,height=91mm}}
  \end{center}
   \vspace*{-7mm}
  \caption{History of deviations of the atmospheric \cft\ concentration over
the last ten millennia relative to the 1850 value  (left-hand scale)
\cite{damon}.  The Medieval Warm (MW), Wolf maximum (W), Sp\"{o}rer
minimum (Sp), Maunder minimum (M) and Suess deviation (Su)  are
indicated.   The data points show the  Earth's magnetic dipole moment over
the same period (right-hand scale)  \cite{damon}. }
  \label{fig_c14_10k_damon}     
  \begin{center}
      \makebox{\epsfig{file=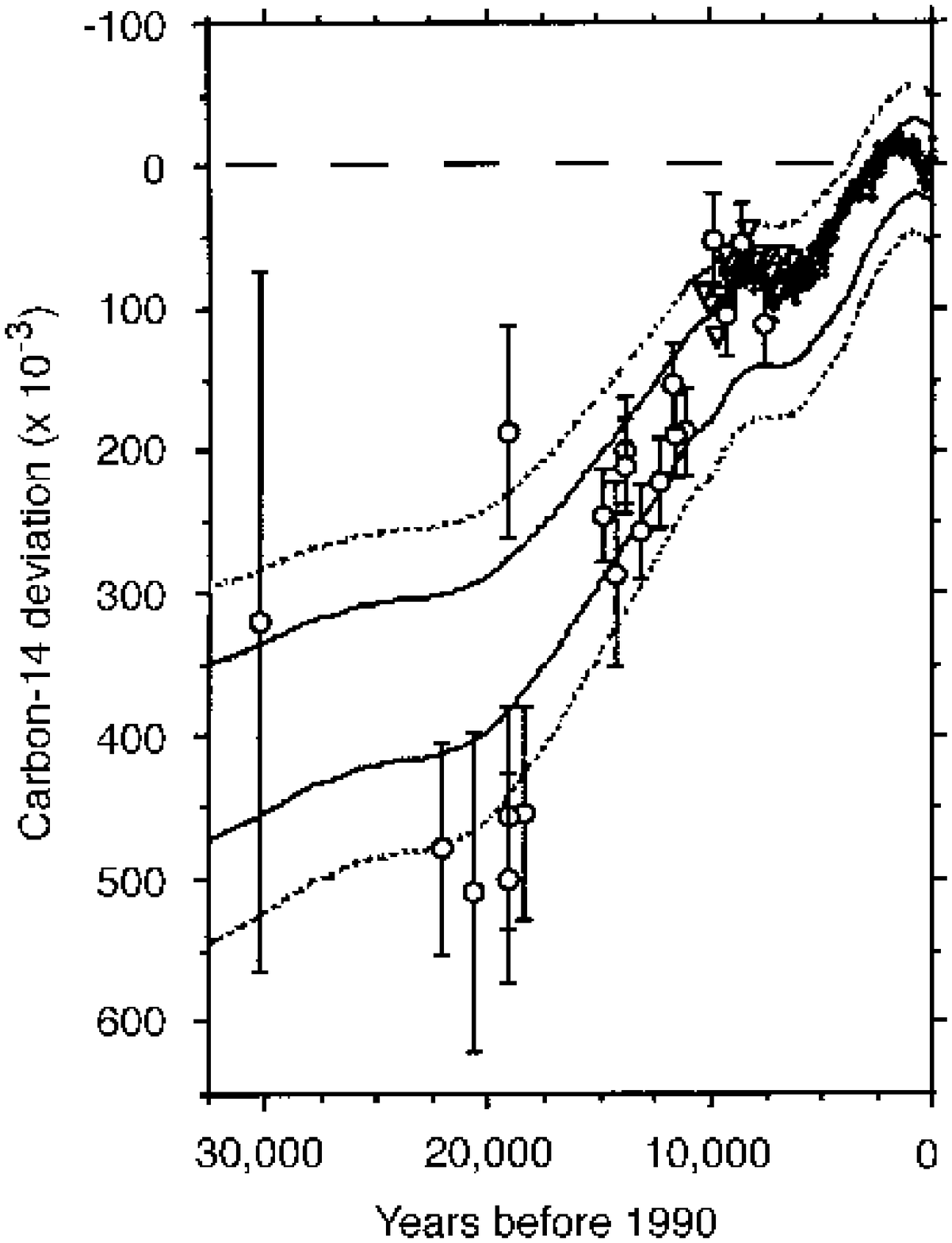,height=88mm}}
  \end{center}
     \vspace*{-7mm}
  \caption{Variation of the atmospheric \cft\ over the last 30 millennia
\cite{bard}.  The two sets of lines correspond to theoretical expectations
based on changes of the Earth's magnetic field; the solid lines assume $\pm
5$\% of the present-day dipole  and the dotted lines, 10\%.}
  \label{fig_c14_30k_bard}  
\end{figure}

\subsubsection{Climatic change since before the last ice age}
\label{sec_ice_age}

The \cft\ record has been extended even earlier; the last 10,000 years is
shown in Fig.~\ref{fig_c14_10k_damon} \cite{lin} and the last 30,000 years
in   Fig. \ref{fig_c14_30k_bard} \cite{bard}).   There are several important
conclusions to be drawn from these data.  Firstly, there have been very
large systematic changes in the cosmic ray flux over this period, namely
about 40\% (where a systematic shift of only 1\% is known to be associated
with a significant climate change, as seen in Fig. \ref{fig_c14_1k_damon}).  
Secondly, the ending of the last great ice age, 17,000--10,000 years ago,
coincided with a sharp decrease of the \cft\ fraction 
(Fig.~\ref{fig_c14_30k_bard}).  Part of this is due to  the reduction in ice
cover and the higher temperatures, thereby increasing the circulation of
\cft-depleted \cotwo\  from the oceans.  An estimated decrease in the
atmospheric \cft\ fraction from these effects is 10\% \cite{damon}. The
measured change (Fig.
\ref{fig_c14_30k_bard}) is substantially larger, suggesting that this period
may have coincided with a decrease of the cosmic flux.   

These large changes in the cosmic ray flux have been caused by changes in
the Earth's magnetic field\footnote{The Earth's field is presently weakening 
(Fig.~\ref{fig_c14_10k_damon}), which may indicate a cooling period
ahead.} and its associated shielding effect.  The geomagnetic field is
independently measured and found to closely track the cosmic ray flux, as
shown in  Figs. \ref{fig_c14_10k_damon} and
\ref{fig_c14_30k_bard}.  This record of changes in the cosmic ray flux,  and
the expected effects on global climate, correlate with the climate records
derived from analyses of cores drilled in the Greenland ice sheet. However
it is important to keep in mind that, even if cosmic rays can indeed affect
cloud formation, it would also require the existence of suitable atmospheric
conditions (humidity, aerosol content, etc.) and that these were probably
quite different during the glacials.

Finally, on an even longer timescale, it is known that the Earth's magnetic
field \emph{reverses} periodically (approximately every million years). 
During these reversals---which may take place fairly rapidly over periods
of less than 1000 years---the Earth's field diminishes to zero and then
re-establishes itself with the opposite polarity.  At such times the Earth will
be subjected to far higher cosmic ray fluxes and the temperatures will be
greatly reduced.  This may perhaps be a factor contributing to the
extinction of the dinosaurs 65 million years ago, which coincided with a
geomagnetic reversal. 

\subsection{Conclusions}

In summary, there are two main conclusions to be drawn on the connection
between cosmic rays and climate change. Firstly, the pattern of systematic
change in the global climate over the past several thousand years follows
the observed changes in the cosmic ray flux over the same period; and it is
consistent with the explanation that a high cosmic ray flux corresponds to
more clouds and a cooler climate, and vice versa. Furthermore, the rise of
about 0.6\degc\ in global temperatures over the last 100 years is consistent
in magnitude and time dependence with the observed changes in cosmic
ray flux---and thereby cloud cover---over the same period.   A clear and
compelling case exists for further study of the effect of cosmic rays on cloud
formation and climate change---especially in order to improve model
calculations of the residual effects due to greenhouse gases from the
burning of fossil fuels. 

Secondly, since the cosmic ray flux is controlled both by the Sun's magnetic
activity (solar wind) and by the Earth's magnetic field, it may eventually be
possible to make long-term (10--1,000 year) predictions of changes in the
Earth's climate by detailed measurements and a  deeper understanding of
the ``magnetic climate'' of the Sun and Earth and its effect on the cosmic ray
flux. In this regard, it will be necessary  to understand how  the Sun
changes its magnetic activity and to understand the progression of the
Earth's magnetic moment.  Perhaps even the longer-term Milankovitch
cycles are partly influencing the global climate through geomagnetic effects;
the Earth's magnetic field is due to  electric currents flowing in the liquid
core and in the magnetosphere, and these may have been influenced by the
periodic gravitational motions of the Milankovitch cycles. 

\section{Cloud formation by charged particles}
\label{sec_cloud_formation}

Cloud droplets in the troposphere\footnote{The troposphere is the lowest
level of the atmosphere and the region where there is enough water vapour
and vertical mixing for clouds to form under suitable conditions.  The depth
of the troposphere is about 6--8 km at the poles, extending to about 17 km
over the equatorial regions; it contains about 75\% of the mass of the
atmosphere.  There is an overall adiabatic lapse rate of temperature in the
troposphere by between 6\degc\ (moist air) and  10\degc\ (dry air)  per
km altitude, reaching a minimum of about -60\degc\ at the boundary with
the stratosphere (the tropopause).  The stratosphere, which extends up to
about 50~km, has a temperature that slowly rises with altitude. This leads
to very little turbulence and vertical mixing and, in consequence, it contains
warm dry air that is largely free of clouds.}  form on suitable aerosols above
about 100 nm size known as cloud condensation nuclei (CCN)
\cite{hobbs,baker}.  The new satellite observations suggest that  the ions or
radicals produced in the atmosphere by cosmic rays may somehow be
affecting the nucleation, growth or activation of atmospheric aerosols, or
affecting the creation of ice particles. 
 An increase of CCN concentration could lead to increased droplet number
concentration, smaller droplet size, less precipitation and therefore longer
cloud lifetime. If cosmic rays can affect ice nuclei then this would also
potentially have a strong influence on clouds since the freezing of
supercooled water droplets reduces their vapour pressure and allows their
growth at the expense of neighbouring liquid droplets. 

A microphysical explanation of cloud formation in the presence of ionising
radiation is presently lacking.  In fact there has been very little research on
this mechanism and present studies do not consider it
\cite{baker}.  Some experiments with highly charged droplets 
\mbox{($10^{-14}$--$10^{-11}$ C)} have, however, shown that they are
10--100 times more efficient at capturing aerosols than uncharged droplets
\cite{barlow}. 

\begin{figure}[htbp]
  \begin{center}
      \makebox{\epsfig{file=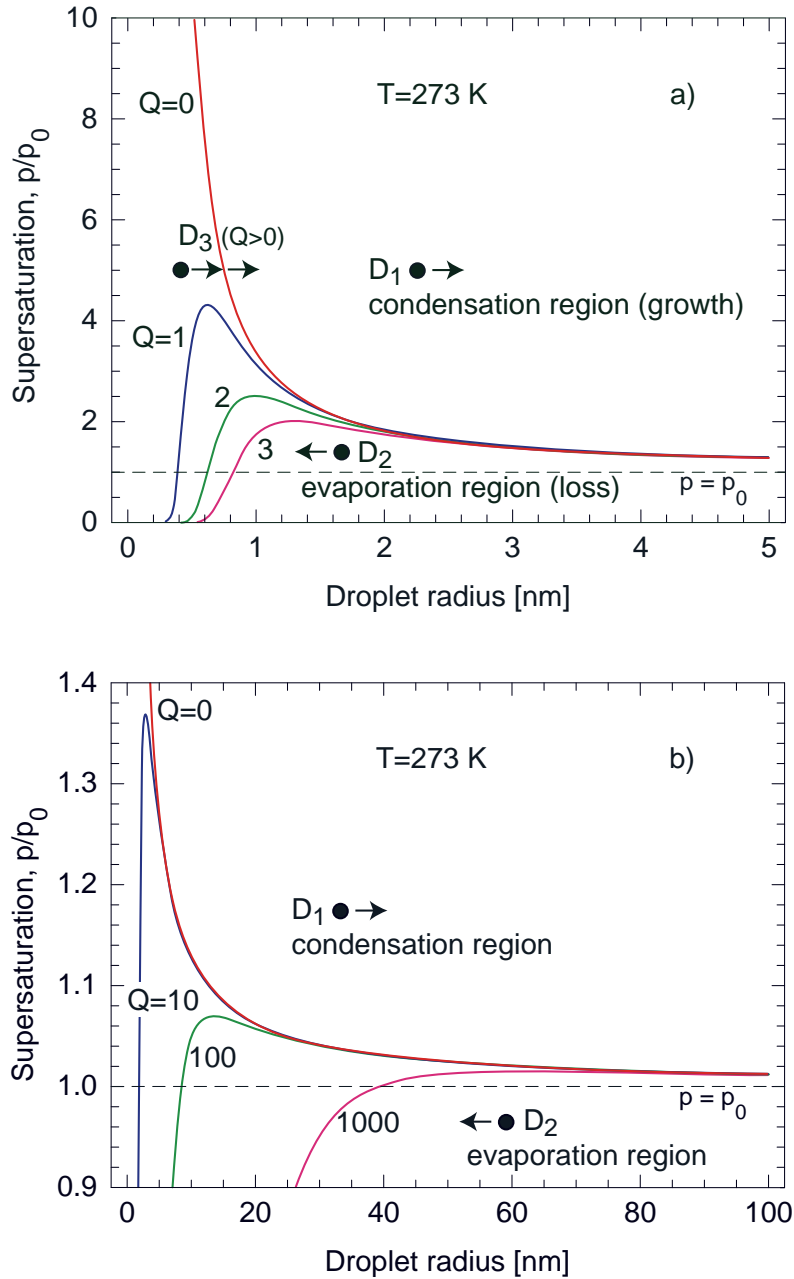,height=170mm}}
  \end{center}
  \caption{Formation of charged droplets at  0\degc\  and various
supersaturations of water vapour, for droplet radii less than a) 5
$\mu$m and b)  100 $\mu$m.  The curves are labelled according to the
droplet electronic charge, Q$e$.  The saturated vapour pressure, p$_0$,
corresponds to equilibrium with a plane water surface.  Droplets formed in
the region above the curves will grow (condensation region) and those
below will shrink (evaporation region).}
  \label{fig_droplet}    
\end{figure}

\subsection{Droplet formation in a conventional cloud chamber}
\label{sec_cloud_chamber}

Certain aspects of how droplets are induced by ionising radiation are known
from the development of cloud chambers, but these largely concern high
supersaturations of water vapour and rapid growth rates.  The principle of
operation of a cloud chamber \cite{segre} can be understood from Fig.
\ref{fig_droplet} which shows the water vapour pressure equilibrium
curves for small droplets:
\[\log _e\left( {{p \over {p_0}}} \right)={M \over {RT\rho }}\left[ {{{2
\gamma } \over r}-\left( {{{Q^2e^2} \over {4\pi \epsilon _0r^2}}\cdot {1
\over {8\pi r^2}}} \right)} \right],\] 
 where $p$ is the vapour pressure, $p_0$ the saturated  vapour pressure at
a plane water surface,  $R$ the gas constant, $T$ the absolute temperature,
$\gamma$ the surface tension, $M$ the molecular weight,
$\rho$ the density, $\epsilon _0$ the permittivity of free space  and $r$ the
radius of the droplet.  The curves divide an upper region of vapour
pressure in which water droplets grow by condensation (e.g. droplet D$_1$)
from a lower region where they shrink by evaporation (e.g. D$_2$).  If a
droplet carries an electric charge $Qe$ then its vapour pressure is reduced
by increased attraction of the polar water molecules.  The growth region is
therefore larger (Fig. \ref{fig_droplet}a) and, in particular, for
supersaturations above about 5, \emph{all} charged droplets will grow and
reach visible size (e.g. D$_3$). The actual starting point for a droplet is not
known.  However, to set the minimum scale we can estimate the volume
occupied by a single water molecule as  
${4 \over 3}\pi r^3={{{M \mathord{\left/ {\vphantom {M
\rho }} \right. \kern-\nulldelimiterspace} \rho }} \over {N_A}}$, where
$N_A$ is Avogadro's constant.  This indicates the effective radius of one
water  molecule is $r \sim 0.2$ nm; and a droplet of radius 0.5 nm contains
about 15 molecules. 

The necessary supersaturation in a cloud chamber is generated by fast
adiabatic expansion.  For an ideal gas and an adiabatic expansion,
\begin{eqnarray*}
  P_1V_1^\gamma      & = & P_2V_2^\gamma, \; \; \; \mbox{and}  \\
  T_1V_1^{\gamma-1} & = & T_2V_2^{\gamma-1},
\end{eqnarray*} 
 where $\gamma$ is the ratio of specific heats
$C_P/C_V$ (about 1.40 for  air and saturated water vapour).  For example,
an expansion ratio $V_2/V_1 =1.3$ results in a temperature drop of about
30\degc\ and a supersaturation of 4.8.  Under these conditions, all droplets
with $Q \geq1$ will grow to diameters of several tens of micrometres in
$\lappeq 0.2$~s.

The curves of Fig. \ref{fig_droplet}a) indicate how  the presence of a small
charge facilitates droplet formation and growth at high supersaturations,
above about 2.  The maximum supersaturations in the troposphere are
typically less than one per cent above unity which, if this were the only
mechanism involved, would require large droplet charges for stable growth,
as indicated in Fig.\,\ref{fig_droplet}b).   Therefore if ionising radiation is
indeed inducing additional cloud formation, it must be somehow affecting
the aerosols or ice nuclei that efficiently seed clouds in the atmosphere. 

\subsection{Ions and droplets in the atmosphere}  \label{sec_ions}

To provide input for the design requirements of the CLOUD detector,  it is
useful to estimate some characteristics  of cosmic-ray ionisation and water
droplets in the troposphere.  Some representative parameters at the upper
and lower boundaries of the troposphere are summarised in
Table~\ref{tab_characteristics}. These regions are well below the 
ionosphere---which starts above the stratosphere at about 80~km
altitude---where atoms and molecules are readily ionised by solar UV and
X-ray photons.  

\begin{table}[h]
  \begin{center}
  \caption{Approximate characteristics of cosmic rays near the top of the
troposphere (10~km altitude) and at sea level. The maximum and minimum
charged cosmic ray fluxes, $\phi$, correspond to the sunspot cycle variation
at geomagnetic latitudes $\gappeq 40^\mathrm{o}$ and for energies
$\gappeq 0.1$ GeV.  The
\# ion pairs~cm$^{-1}, \; n_{ip}$, are the ionising track densities for
minimum ionising particles in air.}
  \label{tab_characteristics}
  \begin{tabular}{| r | r | r | r | r | | r r | r | r |}
 \hline
 \multicolumn{5}{|c||}{Atmospheric characteristics} &
  \multicolumn{4}{|c|}{Cosmic ray characteristics} \\ 
  \hline \hline
 Alt. & Atmos. & Pressure & Density & Mean. & 
\multicolumn{2}{|c|}{Chd. flux, $\phi$} & Ion pairs & Recombination 
\\
     &  depth &     &      & temp. &
\multicolumn{2}{|c|}{[m$^{-2}$sr$^{-1}$s$^{-1}$]} &  /cm, $n_{ip}$ &
coefficient, $\alpha$  \\
  \cline{6-7}
 {[km]} & {[$\lambda_{int}$]}  & {[kN m$^{-2}$]}  & {[kg m$^{-3}$]} &
{[\degc]} &  Max. & Min. & {[cm$^{-1}$]} & {[m$^3$s$^{-1}$]}  \\[0.5ex]
  \hline 
 10 & 3.0 & 26.4  & 0.41 & -50 & 850 & 750 & 23 & $0.64 \times
 10^{-12}$ \\
 0 & 11.5 & 101.3  & 1.22 & 15 & 100 & 97 & 68 & $1.72 \times
 10^{-12}$ \\
  \hline
  \end{tabular}
  \end{center}
\end{table}

The rate of ion-pair production by cosmic rays, 
$ r_{ip}  = \phi \; \Omega \; (100 \ n_{ip})$.  From 
Table~\ref{tab_characteristics} this rate is about  2$\times
10^6$~m$^{-3}$s$^{-1}$ at sea level and $10^7$~m$^{-3}$s$^{-1}$ at 10 km
altitude.   Once created, the charged particles will  recombine.  The electrons
are rapidly captured by electronegative gases, which become negatively
charged as a result.  For example, at 0\degc\ and atmospheric pressure, the
mean electron attachment time is 190 ns in O$_2$ and 140 ns in H$_2$O
vapour.  

The ion mobility is far smaller and consequently so also is the ion
recombination rate:
\begin{eqnarray} 
 {{dn_+} \over {dt}}={{dn_-} \over {dt}}=-\alpha \;n_-\;n_+,
\label{eq_dn/dt}
\end{eqnarray}   where $n_+$ and $n_-$ [m$^{-3}$] are the positive and
negative ion densities and $\alpha$ is the recombination coefficient. 
Equilibrium is reached when the ion production and recombination rates
are equal, i.e. when $r_{ip} = \alpha \; n_+ \; n_-$.  Assuming  $n_+
\sim n_-$ implies the equilibrium ion density,
$n_{eq} = \sqrt {{{r_{ip}} \mathord{\left/ {\vphantom {{r_{ip}} \alpha }}
\right. \kern-\nulldelimiterspace} \alpha }}$. From
Table~\ref{tab_characteristics}, this indicates 
$n_{eq}$ is about  $10^9 \; $m$^{-3}$ at sea level and  
$4\times 10^9 \; $m$^{-3}$  at 10 km.\footnote{For comparison, the droplet
density in clouds is $10^7$--$10^8 \; $m$^{-3}$ over oceans and
$10^8$--$10^9 \; $m$^{-3}$ over land.}   These estimates apply to still air
without aerosols, water droplets, etc. The equilibrium ion density
$n_{eq}
\propto \sqrt  \phi$ so that, for example, the
$\sim$12\% change in cosmic ray flux  at 10 km altitude over a sunspot
cycle produces $\sim$~6\% change in the ion density. 

From equation \ref{eq_dn/dt}, the lifetime of an ion is 
$\tau = 1 / (\alpha \; n_{eq})$.  This implies ion lifetimes in the troposphere
of about 6--9 min.  In still air, the ions drift vertically in the electric field
created by the negatively-charged Earth and the positively-charged
ionosphere.  The field strength is $E \sim 100$ V/m at sea level, producing
an ion drift velocity of about \mbox{0.022 m s$^{-1}$}.  This results in an
ion drift distance of about 12 m before recombination.

In the CLOUD detector, in the absence of a drift field, the ions will diffuse a
projected distance $\sigma_x = \sqrt {2 D t}$ in a time $t$, where
$D$ is the diffusion coefficient.  For water vapour, $D = 1.6 \times 10^{-6}$
m$^2$s$^{-1}$ at conditions corresponding to sea level and
$4.1 \times 10^{-6}$ m$^2$s$^{-1}$ at those corresponding to 10 km
altitude, indicating ion diffusion distances of 4--6 cm before recombination.

In order to be sensitive to long growth times in CLOUD, the droplets must
not be removed by gravitational fall before observation.  This sets a
maximum droplet size for studying long growth times, which can be
estimated as follows. Stoke's formula for the viscous force $F$ on a sphere
of radius $r$ moving at a velocity $v$ is 
\begin{eqnarray} 
 F = - 6 \pi \eta r v, 
\label{eq_stokes}
\end{eqnarray}  where $\eta$ is the dynamic viscosity ($1.7 \times
10^{-5}$  Pa s for air at 0\degc).  This indicates a terminal velocity 
$v_o = (2 r^2 \rho g)/ (9 \eta)$,  or $v_o$ [$\mu$m s$^{-1}$] = 130 $r^2$
[$\mu$m$^2$].   A droplet of radius 1 $\mu$m therefore has a terminal
velocity 130
$\mu$m s$^{-1}$ and falls by about 8 cm during a time of 10~min.  

Very small droplets will be suspended by Brownian motion.  We can
estimate their maximum size as follows. The projected Brownian
displacement is 
$\sigma_x = \sqrt {2 D t}$,  where  
\mbox{$D = \mu k T$}, and $k$ is the Boltzmann constant.  The mobility is 
$\mu = v_{drift}/F  = 1 / (6 \pi \eta r)$ (from equation
\ref{eq_stokes}), where $F$ is the force causing the drift.  Substituting gives
the rms projected displacement of a droplet due to Brownian motion,  
\mbox{$\sigma_x = \sqrt {(k T t) / (3 \pi \eta r)}$}.     Brownian motion will
dominate over gravitational fall when the projected Brownian displacement
is about equal to the droplet radius 
$r$ in the time the droplet would have taken to fall a distance $r$.    This
indicates that Brownian displacement is comparable to gravitational
displacement at a droplet radius of about 0.7 $\mu$m, and  below this
value droplets will remain suspended for extended periods.    

In summary, in order to allow for the possibility of long formation  times in
the atmosphere, the CLOUD detector should have the capability to observe
droplet growth over periods of up to about 10 min, i.e. comparable to the
ion lifetime, and the dimensions of the chamber should allow for a fiducial
volume that is approximately 10 cm away from any boundary.  This
indicates a chamber dimension of about $50
\times 50 \times 50$ cm$^3$ is appropriate.

\begin{figure}
  \begin{center}
      \makebox{\epsfig{file=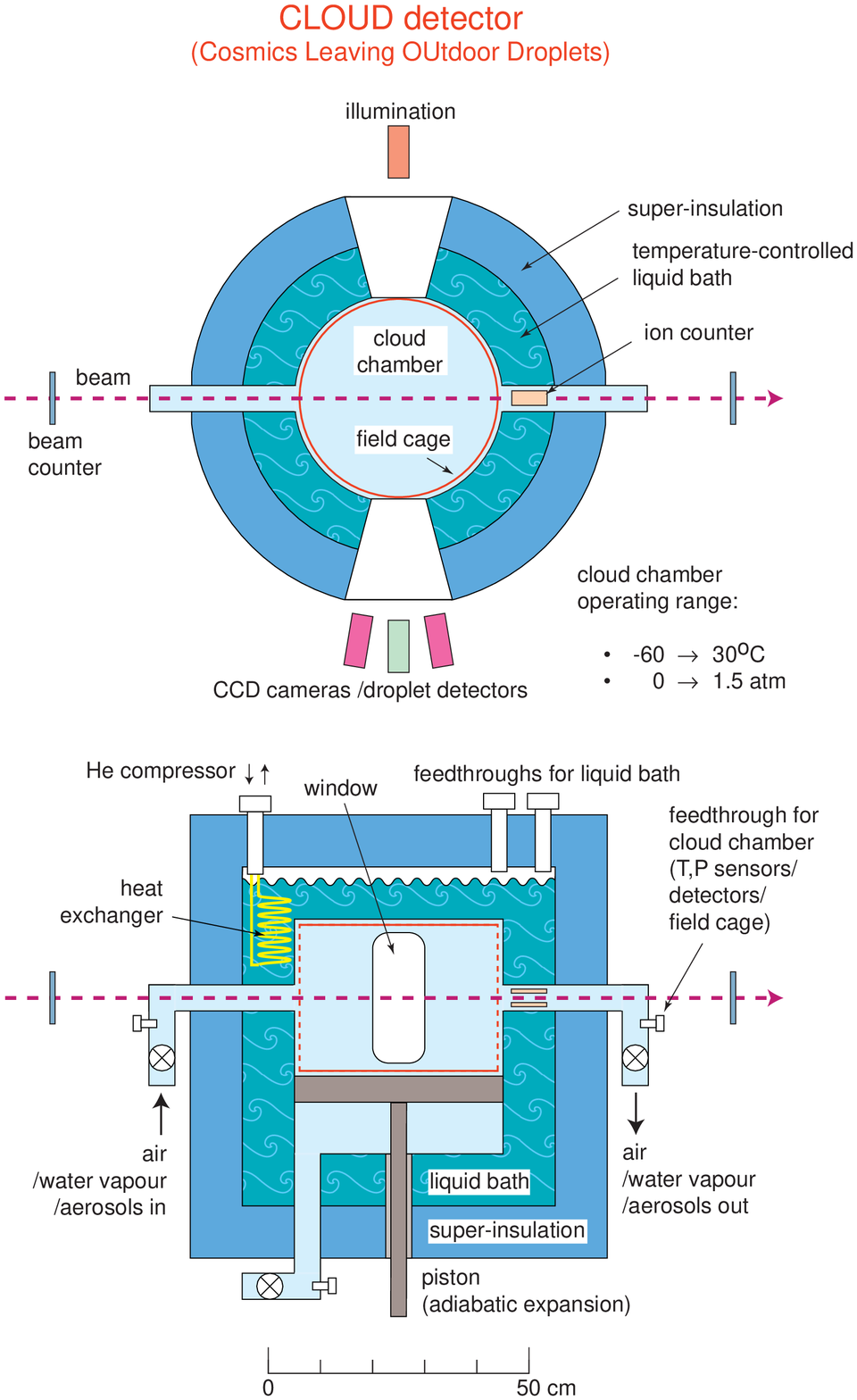,height=220mm}}
  \end{center}
  \caption{Schematic diagram of the CLOUD chamber.}
  \label{fig_detector}    
\end{figure}

\section{The CLOUD experiment}

The basic purpose of the CLOUD detector (Fig.~\ref{fig_detector}) is to
confirm, or otherwise, a  direct link between cosmic rays and cloud
formation by   measuring droplet formation in a controlled test-beam
environment.   The CLOUD detector is essentially a cloud chamber that is
designed to duplicate the gases, water vapour, aerosols and
temperature/pressure conditions found in the Earth's troposphere. 

Since the microphysical mechanisms responsible for cloud formation by
cosmic rays are poorly understood, it is necessary for the experimental tests
to cover a broad range of variables and conditions.  For instance, it is not
yet even known at which \emph{altitude} (i.e. temperature and pressure
conditions) the clouds are being formed.     Also, since the maximum water
vapour  supersaturation in the troposphere is only about one per cent,
droplet growth times may be quite long, as discussed in the previous
section.   The combined effects of ionisation and aerosols/ice particles is
probably important (Section\,\ref{sec_cloud_formation}).  The ionisation of
\emph{existing} water droplets by charged particles requires study;
charged droplets have a lower vapour pressure and will therefore grow at
the expense of uncharged droplets.  It is important to measure the
sensitivity to a range of track ionisation densities, $dE/dx$, and ion pair
concentrations  that duplicates those of cosmic rays.  This indicates the need
for beam particles ranging from protons to heavy ions. Finally, an
investigation should be made of the possibility that  droplets are formed in
relatively ``rare events", such as very highly charged ions or droplets,
perhaps due to heavy nuclei. 

Cloud chamber data under these conditions have never been previously
obtained.  Although C.T.R. Wilson's cloud chamber\footnote{It is interesting 
to note that C.T.R. Wilson had the inspiration for the cloud chamber while
observing meteorological phenomena on the mountain of Ben Nevis in
1894.  (The phenomena were not particle tracks, however, but ``coronas''
around the Sun and glories, where the Sun glows around shadows in the
mist.)} was extensively developed in the first half of this century, it was
optimised for experimental particle physics and operated under conditions
far removed from those of the troposphere. 

\subsection{Detector}

The CLOUD detector (Fig. \ref{fig_detector}) comprises the following main 
components:
\begin{itemize}
\item A temperature-controlled cloud chamber that can be operated in the
temperature range -60\degc\ to 30\degc\ and at pressures from a vacuum
to 1.5 atm. The upper pressure limit is to allow for adiabatic expansions
down to a final pressure of 1 atm. An electrode structure (field cage) in the
cloud chamber provides a simple clearing field for measurements with low
residual ionisation.
\item A surrounding liquid bath enclosed by super-insulation to provide
stable and precise temperature conditions for the cloud chamber.   A
temperature stability of about 0.01~K is required since this is equivalent to
a change of water vapour supersaturation by about 0.05\%. The
investigation of long growth times for the droplets requires that the
operating conditions be held steady for long periods.  The temperature is
adjusted by heat-exchange coils immersed in a suitable  liquid coolant in
the bath and  connected to a He compressor.  The temperature of the
circulating He is controlled by temperature monitors immersed in the
coolant. 
\item A piston to provide adiabatic expansions for investigating rapid
droplet growth times.  Small piston expansions and compressions also
provide a technique for fine control of the water vapour supersaturation in
the cloud chamber.
\item An optical system comprising illumination and stereo microscope
cameras (recuperated from old bubble chamber equipment) which can be
operated with film or CCD readout.  The latter is to allow direct digitization
of the data for analysis. The illumination and optical system must be
capable of detecting and measuring water droplets of sizes down to a
minimum of about 0.5 $\mu$m, i.e. the smallest that can scatter light and
therefore constitute a cloud.  The beam pipe connecting to the cloud
chamber is equipped with further detectors to measure the ion/aerosol
concentration, size distribution and charge. 
\item Temperature and pressure monitors.  
\item A gas system providing air with adjustable amounts of water vapour
and aerosols.  A mass spectrometer is required for physical and chemical
analysis of the incoming and outgoing gas/aerosol mixture.
\item A scintillation counter hodoscope to measure the incident beam. The
beam must be spread over a large area to simulate the quasi-uniform
cosmic radiation. 
\item A DAQ system.
\end{itemize}

\subsection{Experimental goals}

The goals of the proposed study are threefold:
\begin{enumerate}
\item \textbf{Measurements of water droplet (cloud) formation by ionising
radiation}, according to the following variables:
\begin{itemize}
\item Temperature and pressure conditions.
\item Supersaturation of water vapour.
\item Presence of aerosols and trace condensable vapours.
\item Presence of frozen droplets.
\item Presence of existing water droplets.
\item Particle flux and ionisation density (p, heavy ions\ldots).
\item Time for droplet growth. 
\end{itemize}
\item \textbf{Understanding of the microphysical processes} involved in
cloud formation by ionising radiation.  This will involve Monte Carlo
simulation of the processes and a comparison with the CLOUD results.
\item \textbf{Estimation of the global cloud cover} due to cosmic rays, 
using atmospheric data recorded by satellite experiments.
\end{enumerate}

\section{Conclusions}

The observation of an apparent connection between cosmic rays and global
climate offers a unique opportunity for particle physics to make a major
contribution to the problem of global warming---at a relatively modest
cost.  The proposed CLOUD experiment will unambiguously confirm, or
otherwise, a direct link between cosmic radiation and cloud formation, and
allow a quantitative understanding of the physical processes involved.  This,
in turn, would provide an unambiguous and physically explicable
connection between solar and climatic variability---representing a great
advance in the understanding of the relationship between the Earth and its
star.  

Although this experiment primarily addresses atmospheric science, it is
clearly in the domain of particle physics and of CERN's expertise and
experimental facilities.  The CLOUD experiment will have the immediate
impact of  potentially resolving one of the  important unknown effects that
may have so far prevented reliable calculations of global warming from
greenhouse gases---an issue of profound economic and social importance to
the world.   Furthermore, if the link is confirmed, then  it may eventually
become possible to make long-term predictions of systematic changes in the
Earth's climate, by monitoring and predicting the evolution of the cosmic
ray flux and the ``geomagnetic climate'' of the Sun and Earth that modulates
its intensity.  

\section*{Acknowledgements}

I would like to thank Nigel Calder for an inspiring seminar \cite{calder} he
presented on this subject  at CERN, and also thank Horst Wachsmuth who
organised the seminar.  I would also like to acknowledge  advice from
Dietrich Schinzel on the temperature control of the CLOUD detector.


\begin{thebibliography}{99}

\bibitem{ipcc} Intergovernmental Panel on Climate Change (IPCC),
\emph{Climate change 1995: the science of climate change}, eds. J.T.
Houghton et al., WMO and UNEP, Cambridge University Press, Cambridge
(1996).

\bibitem{svensmark97} H. Svensmark and E. Friis-Christensen,
\emph{Variation in cosmic ray flux and global cloud coverage---a missing
link in solar-climate relationships}, Journal of Atmospheric and
Solar-Terrestrial Physics, 59 (1997) 1225.

\bibitem{svensmark98} H. Svensmark, \emph{Influence of cosmic rays on
the Earth's climate}, submitted to Physical Review Letters (1997).

\bibitem{foukal} P.V. Foukal, \emph{The variable Sun}, Scientific American
262 (1990) 26.

\bibitem{eddy} J.A. Eddy, \emph{The Maunder minimum}, Science 192
(1976) 1189.

\bibitem{friis} E. Friis-Christensen and K. Lassen, \emph{Length of the solar
cycle: an indicator of solar activity closely associated with climate}, Science
254 (1991) 698.

\bibitem{willson} R.C. Willson, \emph{Total solar irradiance trend during
solar cycles 21 and 22}, Science 277 (1997) 1963.

\bibitem{ney} E.P. Ney, \emph{Cosmic radiation and the weather}, Nature
183 (1959) 451.

\bibitem{hobbs} P.V. Hobbs, \emph{Aerosol-cloud interactions}, in
\emph{Aerosol-Cloud-Climate Interactions}, International Geophysics
Series, Vol.~54, ed. P.V. Hobbs, Academic Press Inc., San Diego (1993) 33.

\bibitem{baker} M.B. Baker, \emph{Cloud microphysics and climate},
Science 276 (1997) 1072.


\bibitem{hartmann} D.L. Hartmann, \emph{ Radiative effects of clouds on
Earth's climate}, in \emph{Aerosol-Cloud-Climate Interactions}, 
International Geophysics Series, Vol. 54, ed. P.V. Hobbs, Academic Press
Inc., San Diego (1993) 151.

\bibitem{beer} J. Beer, G.M. Raisbeck and F. Yiou, \emph{Time variations of
\bten\ and solar activity}, in
\emph{The Sun in Time}, eds. C.P. Sonett, M.S. Giampapa and
M.S.~Matthews, University of Arizona Press, Tucson (1991).

\bibitem{feynman} J. Feynman and N.U. Crooker, \emph{The solar wind at
the turn of the century}, Nature 275 (1978) 626.

\bibitem{stuiver} M. Stuiver and P.D. Quay, \emph{Changes in atmospheric
carbon-14 attributed to a variable Sun}, Science 207 (1980) 11.

\bibitem{hunten} D.M. Hunten, J.-C. G\'{e}rard and L.M. Fran\c{c}ois,
\emph{The atmosphere's response to solar irradiation}, in
\emph{The Sun in Time}, eds. C.P. Sonett, M.S. Giampapa and
M.S.~Matthews, University of Arizona Press, Tucson (1991).

\bibitem{damon} P.E. Damon and C.P. Sonett, \emph{Solar and terrestrial
components of the atmospheric \cft\ variation spectrum}, in \emph{The
Sun in Time}, eds. C.P. Sonett, M.S. Giampapa and M.S.~Matthews, University
of Arizona Press, Tucson (1991).

\bibitem{lin} Y.C. Lin et al., \emph{Long term modulation of cosmic-ray
intensity and solar activity cycle}, Proc. 14th International Cosmic Ray
Conference, Munich, 3 (1975) 995.

\bibitem{bard} E. Bard et al., \emph{Calibration of the \cft\ timescale over
the past 30,000 years using mass spectrometric U-Th ages from Barbados
corals}, Nature 345 (1990) 405.

\bibitem{barlow} A.K. Barlow and J. Latham, \emph{A laboratory study of
the scavenging of sub-micron aerosol by charged raindrops}, Quarterly
Journal of the Royal Meteorological Society 109 (1983) 763.

\bibitem{segre} E. Segr\`{e}, \emph{Experimental nuclear physics}, Wiley,
New York (1953) 53.

\bibitem {calder} Nigel Calder, \emph{Global warming---blame the Sun},
CERN seminar (December 1997).

\end{thebibliography}
\end{document}